\documentclass{raa}

\usepackage{graphicx,times}
\usepackage{natbib}
\usepackage{amssymb,amsmath}
\usepackage{array}
\usepackage[pagebackref=true]{hyperref}
\bibpunct{(}{)}{;}{a}{}{,}

\newcommand{\Hnot}{H_0}
\newcommand{\HI}{H\,{\sc i}}
\newcommand{\kmsmpc}{\mathrm{km\,s^{-1}\,Mpc^{-1}}}
\newcommand{\lcdm}{$\Lambda$CDM}
\newcolumntype{L}[1]{>{\raggedright\arraybackslash}p{#1}}

\begin{document}

\title{Distance-Ladder Measurements of the Hubble Constant: Recent Progress, Systematics, and Prospects}

\volnopage{Vol.0 (202x) No.0, 000--000}
\setcounter{page}{1}

\author{Xiaodian Chen
   \inst{1,2,3,*}
\and Shu Wang
   \inst{1,3}
}

\institute{National Astronomical Observatories, Chinese Academy of Sciences,
             Beijing 100101, China; {\it chenxiaodian@nao.cas.cn}\\
\and
             Institute for Frontiers in Astronomy and Astrophysics, Beijing Normal University,
             Beijing 102206, China\\
\and
             School of Astronomy and Space Science, University of Chinese Academy of Sciences,
             Beijing 100049, China\\
\vs\no
   {\small Received 202x month day; accepted 202x month day}}

\abstract{The Hubble constant, $\Hnot$, links the nearby distance scale to the present cosmic
expansion rate. Local distance-ladder measurements now reach percent-level precision and remain
more than $5\sigma$ higher than the value inferred from cosmic microwave background (CMB)
observations in base-\lcdm, making the reliability of the local ladder a central issue in the Hubble
tension. We describe the ladder as a covariance network connecting level-0 geometric
anchors, level-1 stellar distance indicators, and level-2 Hubble-flow probes. The Cepheid--Type Ia
supernova (SN Ia) route remains the most precise single local ladder, but independent indicators
including the tip of the red giant branch (TRGB), J-region asymptotic giant branch (JAGB) stars,
Mira variables, surface-brightness fluctuations (SBF), the Tully--Fisher relation, and Type II
supernovae (SNe II) now test shared and method-specific systematics. In a compact
seven-route covariance summary, combining the Cepheid--SN Ia route with three level-1 alternatives (TRGB,
JAGB, and Mira) and three level-2 alternatives (SBF, Tully--Fisher, and SNe II) gives
$\Hnot=73.30\pm0.92~\kmsmpc$, still $5.6\sigma$ above Planck base-\lcdm. JWST has already
tested Cepheid crowding and is making independent TRGB-based $\Hnot$ measurements increasingly
feasible. Over the
next five years, a reliable one-percent local $\Hnot$ requires larger calibrator samples,
cross-validated level-1 zero points, explicit covariance propagation, and AI-assisted,
reproducible, pre-specified selection criteria for distance-indicator measurements.
}

\authorrunning{X. Chen \& S. Wang}
\titlerunning{Distance-Ladder Measurements of the Hubble Constant}
\date{}

\maketitle

\section{Introduction: The Hubble Constant, Distance Ladders, and the Hubble Tension}
\label{sect:intro}

The Hubble constant $\Hnot$ is the present-day value of the Hubble parameter,
$\Hnot\equiv(\dot a/a)_{t_0}$, and sets the absolute scale for extragalactic distances, cosmic
ages, and low-redshift cosmological tests. In the nearby Universe, the basic observable relation is
deceptively simple: recession velocity is compared with distance. Redshifts can be measured with
high precision, but cosmic distances span too wide a range for almost any single indicator to
cover the path from the Milky Way to the Hubble flow. The distance ladder addresses this scale
mismatch by using each indicator where it is best calibrated, from geometric anchors to stellar standard
candles and then to secondary indicators that reach deeper into the Hubble flow, while propagating
statistical and systematic uncertainties from one rung to the next. The historical development of
this problem began with the Cepheid
period-luminosity relation (PLR) discovered by \citet{LeavittPickering1912}, Hubble's use of Cepheids
to establish the extragalactic nature of nearby nebulae \citep{Hubble1926}, and the first
redshift-distance relation \citep{Hubble1929}. By the late twentieth century, however, the local
distance scale still contained an order-unity ambiguity: different calibrations favored
$\Hnot\simeq50~\kmsmpc$ in the long distance scale advocated by Sandage and Tammann, or
$\Hnot\simeq100~\kmsmpc$ in the short distance scale defended by de Vaucouleurs
\citep{SandageTammann1976,deVaucouleurs1982,Tully2024Historical}. That factor-of-two disagreement reflected Cepheid
zero points, extinction, metallicity, galaxy selection, and the calibration of secondary distance
indicators. The Hubble Space Telescope (HST) Key Project was designed to break this deadlock with
a uniform Cepheid-based calibration of several secondary indicators, and it established a landmark
value of $\Hnot=72\pm 8~\kmsmpc$, reducing the historical debate to roughly ten percent
\citep{Freedman2001}.

After the HST Key Project, the field moved from resolving gross disagreements to testing
few-percent and then percent-level systematics. The Carnegie Hubble Program (CHP) used Spitzer
$3.6~\mu$m Cepheid photometry to reduce dust and metallicity sensitivity, obtaining
$\Hnot=74.3\pm2.1\mathrm{(sys)}~\kmsmpc$ and demonstrating how infrared calibration could
sharpen the local ladder \citep{Freedman2012CHP}. By the early 2010s, the distance-ladder question
had become whether these rungs could be tied together without percent-level biases, rather than
whether the ladder could be built at all. The modern crisis began when the Planck
satellite measured the cosmic microwave background (CMB) anisotropy with enough precision to make the early-Universe inference
equally sharp: its first cosmological release already gave
$\Hnot=67.3\pm1.2~\kmsmpc$ in the flat six-parameter \lcdm\ model \citep{Planck2014},
and the final full-mission analysis tightened this to
$\Hnot=67.36\pm0.54~\kmsmpc$ in base \lcdm, with simple one-parameter extensions providing
no preferred solution \citep{Planck2020}. In parallel, the Supernova H0 for the Equation of State
(SH0ES) program built a
Cepheid--Type Ia supernova (SN Ia) ladder tied to multiple geometric anchors
\citep{Riess2016,Riess2019,Riess2021,Riess2022,RiessBreuval2024Local}. Its 2022 analysis found
$\Hnot=73.04\pm1.04~\kmsmpc$ from 42 SNe Ia in 37 Cepheid host galaxies, differing at about
$5\sigma$ from the final Planck \lcdm\ prediction \citep{Riess2022}. This mismatch is commonly
called the Hubble tension \citep{Verde2019,Shah2021Buyer,DiValentino2021,
DiValentino2021CosmoIntertwinedII,Abdalla2022CosmologyIntertwined,
Perivolaropoulos2022Challenges,Verde2023}. Its importance is not only
numerical: if local distance-ladder systematics and early-Universe data analyses both survive
scrutiny, the disagreement would indicate a failure of the minimal six-parameter \lcdm\ mapping
from the CMB to the present-day expansion rate. Proposed new-physics directions include
early dark energy \citep{Poulin2019EDE,KamionkowskiRiess2023}, extra relativistic energy density or dark radiation
\citep{Bernal2016TroubleH0}, non-standard neutrino interactions \citep{Kreisch2020Neutrino},
interacting dark sectors \citep{DiValentino2017IDE}, modified gravity
\citep{DeFelice2020Proca}, and broader pre-recombination changes that reduce the sound horizon or
alter the expansion rate before recombination \citep{Aylor2019SoundHorizon,KnoxMillea2020,
Schoneberg2022H0Olympics}.

This review focuses on the local distance-ladder side of the problem, because before 2030 this
route is likely to remain the most direct arena for improving $\Hnot$. On the early-Universe side,
substantially sharper CMB tests depend on new instruments and data releases. Other
independent late-Universe probes, including baryon acoustic oscillation (BAO)-calibrated inverse ladders
\citep{Alam2021eBOSS,Efstathiou2021H0,DESI2025BAO}, strong lenses
\citep{Treu2022TimeDelay}, and gravitational-wave standard sirens
\citep{Abbott2017StandardSiren,Chen2018Sirens}, provide cross-checks, but their systematics, sample
sizes, and cross-calibrations also still require time to mature. Our goal is to review how local
measurements of $\Hnot$ are
constructed, why the Cepheid--SN Ia ladder remains the highest-precision route, and how independent
indicators such as the tip of the red giant branch (TRGB), J-region asymptotic giant branch (JAGB)
stars, Mira variables, surface-brightness fluctuations (SBF), the Tully--Fisher relation, and Type
II supernovae (SNe II) test the same distance scale \citep{Freedman2021,FreedmanMadore2023Direct}.
Section~\ref{sect:framework} defines the ladder
framework; Section~\ref{sect:cepheid} reviews the Cepheid--SN Ia distance ladder;
Section~\ref{sect:independent} discusses alternative level-1 and level-2 routes together with
method networks;
Section~\ref{sect:systematics} summarizes the error budget; and Section~\ref{sect:future}
outlines the observational and statistical steps needed for a more objective percent-level local
$\Hnot$ measurement.

\section{The Framework of the Modern Cosmic Distance Ladder}
\label{sect:framework}

The local distance ladder transfers an absolute distance scale from nearby geometric measurements
to galaxies in the Hubble flow. We use a level-0, level-1, and level-2 notation. The terminology makes the geometric
foundation explicit and maps directly onto the ``three-step'' or ``three-rung'' Cepheid--SN Ia
ladder used by SH0ES: geometric distances to Cepheids, Cepheid distances to SN Ia host galaxies,
and SNe Ia in the Hubble flow \citep{Riess2022}. The same structure accommodates other routes,
but at different rungs: TRGB, JAGB stars, and Miras can replace the Cepheid level-1 rung,
whereas SBF, Tully--Fisher, and SNe II can replace or supplement the level-2/Hubble-flow
rung once their calibrations and scatter allow \citep{Freedman2019,Blakeslee2021,
DeJaeger2022,Beaton2016CCHP,Freedman2025JWST}.

The level-0 rung consists of geometric anchors. These are distances measured with minimal dependence on
stellar-population or explosion physics: trigonometric parallaxes in the Milky Way, detached
eclipsing binaries in the Large Magellanic Cloud (LMC), and water masers in NGC 4258. The level-0 rung
sets the absolute magnitude, zero point, or PLR of level-1 stellar distance
indicators. Classical Cepheids provide the most widely used level-1 indicator, while TRGB stars,
JAGB stars, Mira variables, and RR Lyrae stars provide independent or complementary stellar
routes. Level-1 measurements use these calibrated stellar standard candles to determine distances to nearby
galaxies, especially galaxies that hosted well-observed SNe Ia in HST or James Webb Space
Telescope (JWST) programs. These
same galaxies then calibrate the level-2 distance indicator, dominated by standardized SNe Ia.
The level-2 rung uses SNe Ia in the smooth Hubble flow to compare distance with redshift and infer
$\Hnot$.

The physical logic of the ladder is to use each method in the range where its calibration is best
understood. Geometric distances provide the absolute scale with few astrophysical assumptions and comparatively explicit
systematic errors, while the objects for which such distances can be measured directly are usually
limited in number and distance. Stellar and extragalactic distance indicators reach far beyond the
geometric anchors, but their standardization depends on stellar evolution, dust, population effects,
photometric calibration, explosion physics, or galaxy scaling relations. A network description
keeps these dependencies visible: independent anchors, independent level-1 indicators, and multiple level-2
samples test different systematics while sharing a common statistical model and covariance
model \citep{Riess2022,Brout2022}.

Much of the present debate can be stated as two quantitative questions that recur below:
whether different level-1 indicators share a common zero point at the sub-percent scale, and
whether the top rung calibrates the Hubble-flow observable without a coherent population, dust, or
velocity-field mismatch.

The level-0 calibration carries the largest leverage over the final scale. A biased geometric
distance sets the absolute scale for every higher level. If a level-0 distance is underestimated, the
calibrated standard candles are inferred to be too faint in absolute magnitude, level-1 galaxy
distances are underestimated, and the final $\Hnot$ is overestimated. To first order, a shift
$\Delta\mu_0$ in the zero-level distance modulus propagates as
\begin{equation}
\Delta \log_{10}\Hnot \simeq -0.2\,\Delta\mu_0 ,
\label{eq:zero_shift}
\end{equation}
so a 0.05 mag underestimate in the anchor scale raises $\Hnot$ by about 2.3\%. This relation
explains why anchor consistency, photometric cross-calibration, and the covariance among anchors
receive as much attention as the statistical precision of large Hubble-flow SN samples.

\subsection{A Compact Ladder Formalism}
\label{subsect:formalism}

The mathematical form of the ladder is clearest when written in the same order as the
observations proceed, from the nearest geometric calibrators to the Hubble-flow sample. Here we
use the Cepheid--SN Ia route as the example. Other level-1 indicators, such as TRGB, JAGB stars,
Mira variables, or RR Lyrae stars, can replace Cepheids in the first two steps, and other
level-2 indicators can replace or supplement SNe Ia when their calibrations and scatter allow.
For compactness, the same host index $i$ is used below at different steps of the ladder. Depending
on context, $i$ can denote a level-0 geometric anchor, a level-1 nearby SN Ia calibrator host, or a
level-2 Hubble-flow SN Ia host.

The starting point is the extinction-corrected distance modulus,
\begin{equation}
\mu_0 = m_0 - M = 5\log_{10}\left(\frac{D_L}{\mathrm{Mpc}}\right)+25 ,
\label{eq:distance_modulus}
\end{equation}
where $m_0$ is the apparent magnitude corrected for foreground and host extinction, $M$ is the
absolute magnitude, and $D_L$ is the luminosity distance in Mpc. At level-0, geometric distances
to Milky Way Cepheids, LMC Cepheids, and NGC 4258 Cepheids convert observed Cepheid
magnitudes into absolute magnitudes. For a Cepheid $j$ in host galaxy $i$, with period
$P_{i,j}$ and metallicity indicator $[{\rm O/H}]_{i,j}$, the near-infrared (NIR) Wesenheit form of the
Leavitt law can be written as
\begin{equation}
m^W_{H,i,j} = \mu_{0,i} + M^W_{H,1}
             + \beta_W\left(\log P_{i,j}-1\right)
             + \gamma_W[{\rm O/H}]_{i,j},
\label{eq:cepheid_pl}
\end{equation}
following the compact notation used by \citet{Riess2022}. Here $M^W_{H,1}$ is the absolute
Wesenheit magnitude of a 10-day Cepheid at the reference metallicity, $\beta_W$ is the period
slope, and $\gamma_W$ is the metallicity term. In a level-0 anchor system, $\mu_{0,i}$ is supplied by geometry,
so
Equation~(\ref{eq:cepheid_pl}) establishes the Cepheid PLR on an
absolute scale.

The next step applies this calibrated PLR to Cepheids in nearby galaxies that hosted well-observed
SNe Ia. Their Cepheid periods and apparent magnitudes are measured with HST or JWST, and
Equation~(\ref{eq:cepheid_pl}) is solved for the host-galaxy distance modulus $\mu_{0,i}$. This
level-1 step converts a stellar standard candle into the distance of a SN Ia calibrator galaxy.

The calibrated host-galaxy distance then sets the absolute magnitude of the SN Ia in that galaxy.
For a standardized SN Ia in host $i$,
\begin{equation}
m^0_{B,i} = \mu_{0,i} + M^0_B ,
\label{eq:sn_calibrator}
\end{equation}
where $m^0_B$ is the standardized rest-frame $B$-band peak magnitude and $M^0_B$ is the fiducial
SN Ia absolute magnitude. A sample of nearby SN Ia calibrators determines $M^0_B$ and transfers
the stellar distance scale onto SNe Ia.

\begin{figure*}
\centering
\includegraphics[width=\textwidth]{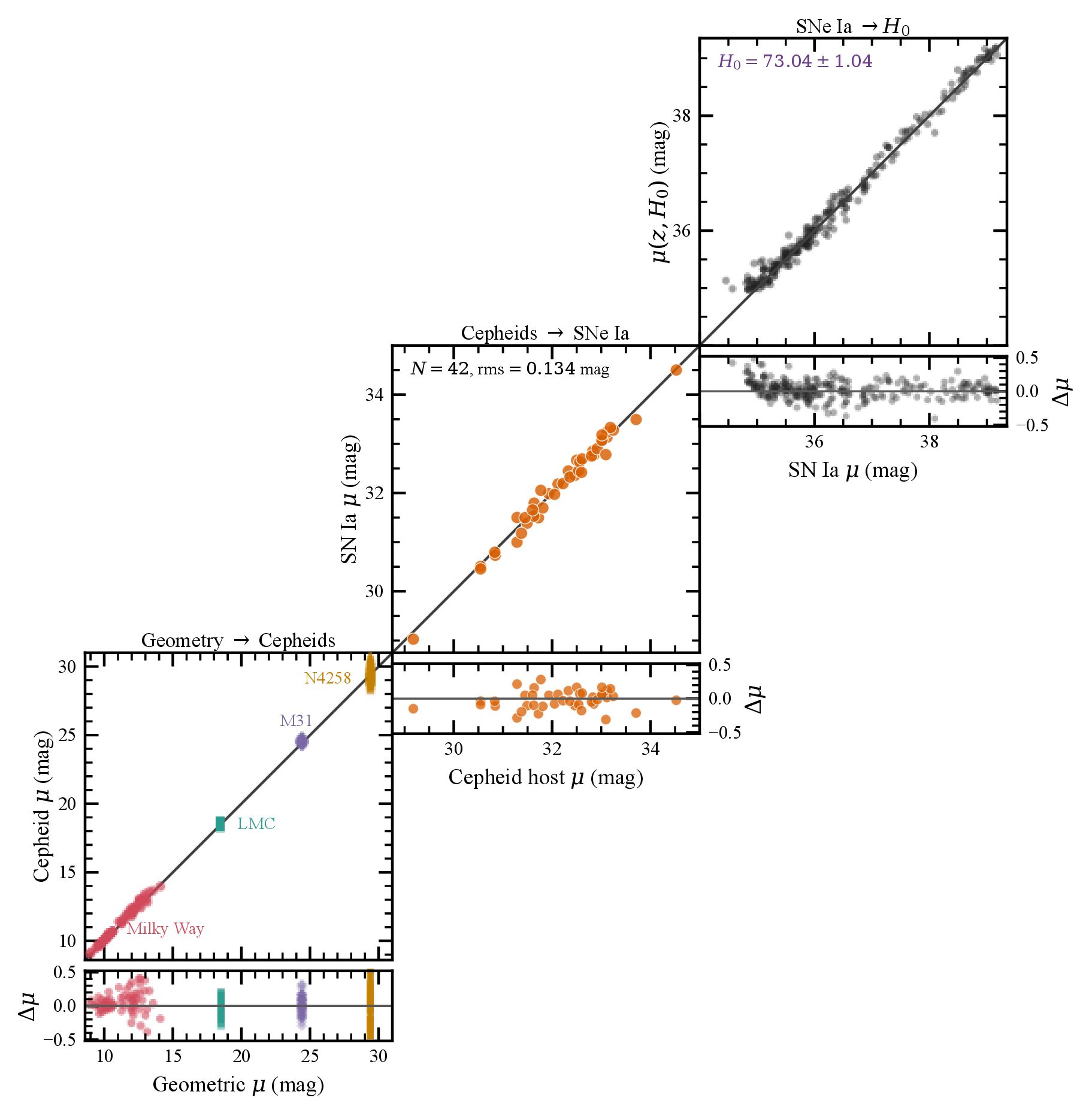}
\caption{Example of the Cepheid--SN Ia distance ladder based on data from \citet{Riess2022}.
Left: geometric distance moduli for
Milky Way Gaia Cepheids, the LMC, M31, and NGC~4258 compared with Cepheid PLR-inferred moduli.
Middle: Cepheid calibrator-host moduli compared with SN Ia moduli for the 42 calibrator SNe Ia.
Right: Hubble-flow SN Ia moduli compared with redshift moduli evaluated at the fitted $\Hnot$.
The lower panels show residuals in distance modulus. This distance-modulus comparison illustrates
how the distance ladder is assembled rung by rung.}
\label{fig:riess2022_ladder_matrix}
\end{figure*}

The final step uses standardized SNe Ia in the Hubble flow. These objects share the calibrated
absolute magnitude $M^0_B$, reach far beyond the local velocity field, and determine the intercept
of the magnitude-redshift relation. In the low-redshift limit,
\begin{equation}
a_B \simeq \log_{10}(cz_i) - 0.2\,m^0_{B,i} ,
\label{eq:ab_simple}
\end{equation}
which is adequate only when higher-order terms in the luminosity-distance expansion are
negligible. For the redshift range commonly used to measure the local Hubble-flow intercept,
$0.023\lesssim z\lesssim0.15$ in SH0ES/Pantheon+ analyses, the cosmographic expansion of the
luminosity distance is retained \citep{Visser2005,Riess2022,Brout2022}:
\begin{equation}
D_L(z_i)=\frac{cz_i}{\Hnot}\left[1+\frac{1-q_0}{2}z_i
-\frac{1-q_0-3q_0^2+j_0}{6}z_i^2+O(z_i^3)\right],
\label{eq:cosmographic_dl}
\end{equation}
where $q_0$ and $j_0$ are the present-day deceleration and jerk parameters. Equivalently, the
intercept fitted from Hubble-flow SNe Ia can be written as
\begin{equation}
a_B = \log_{10}\left\{cz_i\left[1+\frac{1-q_0}{2}z_i
-\frac{1-q_0-3q_0^2+j_0}{6}z_i^2+O(z_i^3)\right]\right\}
-0.2\,m^0_{B,i}.
\label{eq:ab_cosmographic}
\end{equation}
The Hubble constant then follows from
\begin{equation}
\log_{10}\Hnot = 0.2\,M^0_B + a_B + 5 .
\label{eq:h0_formula}
\end{equation}
Equations~(\ref{eq:distance_modulus})--(\ref{eq:h0_formula}) summarize the near-to-far logic of
the ladder: geometric distances establish the Cepheid PLR, the Cepheid PLR measures distances to
SN Ia host galaxies, those distances calibrate the SN Ia luminosity, and Hubble-flow SNe Ia then
convert the calibrated luminosity scale into $\Hnot$.
Figure~\ref{fig:riess2022_ladder_matrix} visualizes the same sequence by comparing the distance
modulus carried from one rung with the distance modulus inferred at the next rung.

Current measurements evaluate these relations as a single global, covariance-aware
fit from the level-0 anchors to $\Hnot$. The observables are collected into a data vector
$\boldsymbol{y}$, the dependence on parameters such as anchor offsets, PLR coefficients,
host-galaxy distances, $M^0_B$, and $a_B$ is encoded in a design matrix $\boldsymbol{L}$, and
statistical plus systematic uncertainties enter through a covariance matrix $\boldsymbol{C}$. For
the linear part of the problem this is a generalized least-squares calculation,
\begin{equation}
\chi^2=(\boldsymbol{y}-\boldsymbol{L}\boldsymbol{q})^{T}
\boldsymbol{C}^{-1}(\boldsymbol{y}-\boldsymbol{L}\boldsymbol{q}),
\label{eq:global_fit}
\end{equation}
while Markov Chain Monte Carlo or nested-sampling analyses are used to sample the corresponding
posterior, test the Gaussian approximation, and propagate priors or non-linear cosmological
parameters \citep{Riess2022,Brout2022}. In this formulation, the distance ladder is a set of shared
terms: common anchors, PLR parameters, photometric calibration,
metallicity terms, SN standardization, redshift corrections, and Hubble-flow intercepts all carry
covariances that are propagated coherently to the final $\Hnot$.

\subsection{Geometric Anchors}
\label{subsect:anchors}

The most widely used geometric anchors now play complementary roles and have reached the
precision needed to test one another. In the Milky Way, trigonometric parallaxes now come
primarily from Gaia, but Cepheid applications require an explicit parallax zero-point model. For
Gaia Data Release 2 (DR2), the astrometric solution showed a global quasar-based parallax offset of about
$-0.029$ mas \citep{Lindegren2018}, while Cepheid-based tests found a larger, sample-dependent
offset of order $-0.046$ mas \citep{Riess2018}. Gaia Early Data Release 3 (EDR3) reduced the problem by providing a
color-, magnitude-, and sky-position-dependent correction; after applying the official correction,
the bright Cepheids used by SH0ES showed a residual offset of about $-14\pm6~\mu$as
\citep{Lindegren2021,Riess2021}. In the same analysis, Gaia EDR3 parallaxes of Milky Way
Cepheids gave a near-infrared Wesenheit Cepheid zero point
$M^W_{H,1}=-5.915\pm0.022$ mag; in the later SH0ES global fit, using the same optimized PLR
parameters for all Cepheids, this Milky Way EDR3 constraint is quoted as
$M^W_{H,1,{\rm Gaia}}=-5.903$ with $\sigma_{\rm Gaia}=0.024$ mag \citep{Riess2021,Riess2022}.
The $\sigma=6~\mu$as residual parallax-offset uncertainty alone corresponds to about 0.9\% in
distance for the mean parallax of the Milky Way Cepheid sample \citep{Riess2022}. Independent
checks with eclipsing binaries, including the W UMa sample of \citet{Ren2021}, also show that
EDR3 reaches percent-level distance calibration only when the source-dependent correction is applied.

The LMC has long served as the first major extragalactic step in cosmic distance measurements
because it lies beyond the Milky Way while remaining rich in Cepheids and other resolved stellar
populations. For decades, LMC distance determinations were numerous and sometimes discordant;
the HST Key Project adopted $\mu_{\rm LMC}=18.50\pm0.10$ mag, corresponding to an uncertainty
of roughly 5\% in distance, and later reviews continued to emphasize the correlated systematics
among published LMC distances \citep{Freedman2001,DeGrijs2014}. Detached eclipsing binaries
made the LMC anchor much sharper. A measurement based on eight late-type eclipsing systems gave
$\mu_{\rm LMC}=18.493\pm0.008_{\rm stat}\pm0.047_{\rm syst}$ mag, a 2.2\% distance
\citep{Pietrzynski2013}. The later analysis of 20 detached eclipsing binaries reached 1\% precision,
$\mu_{\rm LMC}=18.477\pm0.004_{\rm stat}\pm0.026_{\rm syst}$ mag, or
$D=49.59\pm0.09_{\rm stat}\pm0.54_{\rm syst}$ kpc \citep{Pietrzynski2019}.

NGC 4258 supplies an external geometric anchor through water-maser orbits in its nuclear disk. The
first geometric maser distance was $7.2\pm0.3$ Mpc \citep{Herrnstein1999}; subsequent disk
modeling gave $7.60\pm0.17_{\rm stat}\pm0.15_{\rm syst}$ Mpc \citep{Humphreys2013}, and the
current commonly used value is $7.576\pm0.082_{\rm stat}\pm0.076_{\rm syst}$ Mpc, a 1.5\%
distance \citep{Reid2019}.

These anchors are precise and mutually independent. In the global SH0ES solution, the three main
geometric anchors are combined through a covariance-aware fit, approximately an inverse-variance
combination of independent zero point information with additional covariance from PLR slope,
metallicity, and photometric terms. The resulting Cepheid Wesenheit zero point is
$M^W_{H,1}=-5.894\pm0.017$ mag, or about 0.8\% in distance \citep{Riess2022}. The agreement of
the three anchors tests the level-0 scale itself: they determine the same
Cepheid zero point through unrelated geometric measurements before that scale is transmitted to
the rest of the ladder. Gaia parallaxes are expected to carry increasing weight in this level-0
calibration. Gaia Data Release 4 (DR4) is planned to use a longer astrometric time span than EDR3 and an expanded
set of astrometric products; if its bright-star zero-point calibration improves as expected, the
Milky Way parallax anchor becomes a higher-weight, internally checked route to the
Cepheid PLR zero point.

\subsection{Physical Distance Indicators}
\label{subsect:indicators}

Level-1 and level-2 measurements are physical distance measurements. In the terminology used
here, level-1 indicators are calibrated by level-0 anchors and determine distances to nearby
galaxies, while level-2 indicators carry those calibrations into the Hubble flow. They rely on
empirical standardization relations whose zero points are fixed either by level-0 geometric
distances or by the preceding level of the ladder. For pulsating stars the basic calibration is a
period-luminosity-metallicity (PLZ) relation,
\begin{equation}
  M_\lambda = M_{\lambda,0}+\beta_\lambda(\log P-P_0)+\gamma_\lambda{\rm [Fe/H]} ,
\end{equation}
or a multi-band Wesenheit form designed to reduce reddening sensitivity \citep{Madore1982,
Ripepi2022,Riess2022}. Here $M_\lambda$ is the absolute magnitude in passband $\lambda$,
$P$ is the pulsation period, usually in days, $P_0$ is a reference value of $\log P$,
$\beta_\lambda$ is the period slope, $\gamma_\lambda$ is the metallicity coefficient, and
$M_{\lambda,0}$ is the zero point at the reference period and metallicity. For candles without a
period-based calibration, such as the TRGB or JAGB stars, the calibration is an absolute magnitude or a
luminosity-function feature. The relevant uncertainties include the zero point, the slope or edge
definition, metallicity and population terms, extinction, crowding, and photometric
cross-calibration.

Classical Cepheids are young, intermediate- to high-mass, core-helium-burning pulsators crossing
the instability strip. Their Leavitt law \citep{LeavittPickering1912} is mainly set by the coupling of
the period--mean-density relation with the stellar mass--luminosity relation, with temperature,
metallicity, helium abundance, convection, and mass loss producing wavelength-dependent
corrections \citep{Anderson2016,DeSomma2020,DeSomma2021}. Fundamental-mode Cepheids used
in distance-scale work typically have periods from a few days to tens of days; the long-period
Cepheids used in SN Ia hosts often have $P\gtrsim10$ d, with absolute magnitudes ranging
from roughly $M_V\sim -3$ to $-6$ mag and still brighter values in the near-infrared
\citep{DeSomma2020,DeSomma2021,Ripepi2022}.
Their high luminosities and periodic light curves allow direct identification in nearby star-forming
galaxies; HST extended Cepheid searches from $\sim25$ Mpc with the Wide Field and Planetary
Camera 2 (WFPC2) to $\sim40$ Mpc with the Advanced Camera for Surveys (ACS)
and Wide Field Camera 3 (WFC3) \citep{Riess2019}, and SH0ES now measures Cepheids in SN Ia hosts at distances of
$\sim40$--50 Mpc \citep{Riess2022}.

Recent PLR work is dominated by Gaia parallaxes, Optical Gravitational Lensing Experiment
(OGLE) Magellanic Cloud samples, open-cluster
Cepheids, and infrared photometry. In well-controlled near- and mid-infrared samples, the PLR
scatter is typically $\sim0.07$--0.10 mag, corresponding to a single-Cepheid distance precision of
about 3--5\% before averaging many variables in one host. Current geometric and cluster-based
zero-point calibrations reach the percent level; examples include the Gaia EDR3/HST calibration,
Gaia-band period-Wesenheit-metallicity relations, and the 0.9\% open-cluster Cepheid luminosity
scale. In the notation of the PLZ equation, the recent SH0ES global fit gives
$\beta_W=-3.299\pm0.015$ mag dex$^{-1}$ in period,
$\gamma_W=-0.217\pm0.046$ mag dex$^{-1}$ in metallicity, and
$M^W_{H,1}=-5.894\pm0.017$ mag
\citep{Wang2018Cepheid,Chown2020,Riess2021,Ripepi2022,CruzReyes2023,Riess2022}. The leading
Cepheid systematics are the PLR zero point, metallicity dependence, reddening law, crowding and
blending in star-forming disks, and the period distribution mismatch between Milky Way, LMC,
NGC 4258, and SN Ia host samples \citep{Breuval2022,Bhardwaj2024}.

The TRGB traces the luminosity cutoff reached by old, low-mass red giants just before the helium
flash. In the $I$ band the bolometric correction and color dependence partially compensate, giving
a sharp edge in the luminosity function \citep{Lee1993,Rizzi2007}. For old, metal-poor red giant branch (RGB)
populations, approximately $-2.2\lesssim{\rm [Fe/H]}\lesssim-0.7$ dex in the conventional
logarithmic abundance scale or $1.5\lesssim(V-I)_0\lesssim2.0$ mag in color, the $I$-band
TRGB luminosity depends only weakly on
metallicity. The empirical color correction of \citet{Rizzi2007}, for example, has a slope of
0.217 mag per mag in $(V-I)_0$, so the correction across the usual blue-halo selection is only of
order 0.05--0.10 mag. The Carnegie-Chicago Hubble Program (CCHP) LMC-based
calibration used $M_I({\rm TRGB})=-4.049\pm0.022\mathrm{(stat)}\pm0.039\mathrm{(sys)}$ mag
\citep{Freedman2019,Pietrzynski2019}. TRGB measurements are usually made in galaxy halos, where
dust and crowding are reduced, but their accuracy depends on edge-detection methodology,
color/metallicity correction, AGB contamination, photometric depth, and whether the selected halo
field fairly samples the old stellar population \citep{Rizzi2007,JangLee2017,Freedman2019}.

JAGB stars are carbon-rich asymptotic giant branch (AGB) stars whose near-infrared luminosity function has a relatively
stable mode. The stars are bright in the near-infrared and occupy a
different stellar population from Cepheids and TRGB stars \citep{MadoreFreedman2020}. JWST
observations show a JAGB luminosity-function width of about 0.32 mag in early SN Ia host tests
\citep{Lee2024JAGB}. In the expanded JWST sample, JAGB distances agree with HST Cepheid
distances by $-0.03\pm0.02\mathrm{(stat)}\pm0.05\mathrm{(sys)}$ mag on average, while the
dominant calibration issue is the $0.11\pm0.022$ mag field-to-field difference between the inner
and outer NGC 4258 calibration fields \citep{Li2025JAGB}. This difference is larger than the
0.032 mag uncertainty of the maser distance to NGC 4258 and appears, in related form, in
independent NGC 4258 fields. It points to luminosity-function shape and field selection as the
current limiting step of the JAGB anchor; using non-mode statistics can reduce the field difference,
but anchor-field variance remains an explicit systematic \citep{Li2025JAGB}. Here, non-mode statistics refers to
estimators of the JAGB luminosity-function location based on quantities such as the mean, median,
quantiles, or a fitted model center rather than the histogram peak itself; these alternatives can
be less sensitive to binning, smoothing, and small changes in field selection. The relevant systematics are the
luminosity-function selection window, age and metallicity effects in carbon-star formation,
reddening, and anchor-field variance \citep{Lee2024JAGB,Li2025JAGB}.

Mira variables are long-period AGB pulsators with tight near-infrared PLRs, especially after
period cuts remove heavily dust-enshrouded or hot-bottom-burning objects. NGC 4258 provides a
geometric anchor for the Mira PLR \citep{Huang2018}, and HST observations have applied the
method to SN Ia hosts including NGC 1559 and M101 \citep{Huang2020,Huang2024}. Selected
O-rich Miras have near-infrared PLR scatter of about 0.12 mag in local calibrating samples, while
HST applications in SN Ia hosts show broader observed relations, for example about 0.24 mag in
M101. The current Mira route has lower precision than the best Cepheid and TRGB ladders, with
the current two-host Mira--SN Ia ladder giving
$\Hnot=72.37\pm2.97~\kmsmpc$, or about 4.1\% total precision; its independent value comes from
older stellar populations and infrared time-domain information. The leading systematics are
circumstellar dust, period selection, metallicity, sparse phase sampling, and the connection between
O-rich and C-rich AGB subsamples.

RR Lyrae stars are old, low-mass horizontal-branch pulsators. They are fainter than Cepheids and
mostly calibrate the nearby distance scale, dwarf galaxies, globular clusters, and stellar
halo populations, but their near-infrared PLZ relations are physically well motivated and weakly
affected by extinction \citep{Catelan2004,Muraveva2018}. Gaia parallaxes and HST photometry now
give zero-point tests at the few hundredths of a magnitude level, with metallicity entering both the
optical $M_V$--[Fe/H] relation and infrared PLZ relations \citep{Muraveva2018,Neeley2019}. In
well-sampled infrared data, RR Lyrae PLZ or mid-infrared PLZ relations have dispersions of
roughly 0.05--0.08 mag, close to 2--4\% in distance for a single star
\citep{Neeley2019,Mullen2023}. The metallicity term is informative but observationally expensive:
large extragalactic RR Lyrae samples require spectroscopy or well-calibrated multi-band
light-curve metallicities, which is costly for faint variables and often sets the observational limit on
PLZ applications. Double-mode RR Lyrae (RRd)
stars, which pulsate simultaneously in the fundamental and first
overtone modes, add an asteroseismic constraint through the period ratio. Recent work shows that
RRd stars can provide distances and metallicities with a zero-point uncertainty near
0.022 mag, making them promising anchors for old populations and Local Group systems
\citep{Chen2023RRd}. Their current limitation is sample size and the need for homogeneous
multi-band light curves.

Other short-period variables can become auxiliary level-1 indicators as time-domain surveys
grow. W~UMa-type contact eclipsing binaries obey a period-color-luminosity relation, with the
recent calibrations moving from optical period-color-luminosity relations to multi-band PLZ
relations \citep{Chen2018WUMa,Li2025WUMa}. The newest Gaia-based calibration reaches a
zero-point precision of about 0.3\% and a minimum PLZ dispersion of
$\sim0.13$--0.14 mag \citep{Li2025WUMa}. Because W~UMa systems occur at a level of order
$10^{-3}$ among stars, modern time-domain surveys can deliver very large samples, making them
promising future high-precision distance indicators if binary evolution, third light, reddening,
metallicity, and sample contamination are controlled. High-amplitude $\delta$~Scuti and
SX Phoenicis stars obey PLRs and are now being recalibrated with Gaia parallaxes. Recent
near-infrared work reports a Gaia parallax zero point of $35\pm2~\mu$as and a PLR zero-point
precision of about 0.9\%, while double-mode $\delta$~Scuti stars provide an additional way to
reduce mode-identification scatter and test the PLR zero point
\citep{Li2025DeltaScuti,Jia2025DeltaScuti}. These stars are much fainter than Cepheids and are
best suited for independent checks in
clusters, nearby galaxies, and old or intermediate-age systems once Gaia, the Vera C. Rubin
Observatory's Legacy Survey of Space and Time (Rubin/LSST), the Nancy Grace Roman Space Telescope
(Roman), and JWST time-domain samples become larger.

Level-2 indicators have to reach beyond the very local velocity field. SNe Ia dominate this level
because their standardized luminosities are bright enough to be measured deep into the smooth
Hubble flow, where peculiar velocities contribute a smaller fraction of the recession velocity.
Physically, normal SNe Ia are thermonuclear explosions of carbon--oxygen white dwarfs in binary
systems; their optical light curves are powered mainly by the radioactive chain
$^{56}{\rm Ni}\rightarrow{}^{56}{\rm Co}\rightarrow{}^{56}{\rm Fe}$, which ties the peak luminosity
to the synthesized $^{56}$Ni mass \citep{Arnett1982,Maoz2014}. This
common explosion channel makes SNe Ia approximate standard candles, but it does not make their
absolute magnitudes identical. Hotter, more luminous events decline more slowly, while
differences in intrinsic color and host-galaxy dust make redder events fainter in the optical. The
empirical width--luminosity relation \citep{Phillips1993} and its physical interpretation in terms
of radioactive heating, diffusion time, opacity, and color evolution \citep{KasenWoosley2007}
provide the basis for standardization.

Modern cosmological analyses usually implement this standardization with a Tripp-style
relation \citep{Tripp1998,Guy2007}. In a compact form,
\begin{equation}
  m_B^0 = m_B+\alpha_{\rm SN}x_1-\beta_{\rm SN}c+\Delta_{\rm host}+\Delta_{\rm bias},
\end{equation}
where $m_B$ is the fitted rest-frame $B$-band peak magnitude, $x_1$ is the light-curve stretch or
shape parameter, $c$ is the color parameter, $\alpha_{\rm SN}$ and $\beta_{\rm SN}$ are empirical
standardization coefficients, $\Delta_{\rm host}$ accounts for correlations with host-galaxy
properties such as stellar mass, and $\Delta_{\rm bias}$ corrects selection and measurement biases
\citep{Guy2007,Brout2022}. Pantheon+ contains 1701 light curves of 1550 distinct SNe Ia
over $0.001<z<2.26$, while the local SH0ES/Pantheon+ intercept uses 277 Hubble-flow SNe Ia
over $0.023<z<0.15$ and 42 SNe Ia in Cepheid-calibrator hosts \citep{Brout2022,Riess2022}.
The standardized absolute magnitude in the SH0ES/Pantheon+ calibration is
$M^0_B=-19.253\pm0.027$ mag, with a Hubble-flow dispersion of about 0.135 mag
\citep{Riess2022}. The main systematics are color-law calibration, host-galaxy correlations,
survey cross-calibration, selection bias, peculiar-velocity corrections, and environmental
differences between calibrator and Hubble-flow samples.

SBF measures the pixel-to-pixel variance of unresolved stellar populations. The fluctuation
luminosity depends on the second moment of the stellar luminosity function and is calibrated as a
function of stellar-population color \citep{TonrySchneider1988,Blakeslee2021}. It is most effective
for early-type galaxies: a single HST orbit can measure near-infrared SBF distances beyond
$\sim80$ Mpc, individual-galaxy precision is typically $\lesssim5\%$, and HST SBF distances have
already been measured for more than 370 galaxies, including more than 220 in the WFC3 infrared
channel (WFC3/IR) $F110W$ band
\citep{Blakeslee2021,Jensen2025SBF}. The recent JWST TRGB--SBF program calibrates HST SBF
with JWST TRGB distances tied to NGC 4258; the current analysis uses eight galaxies with direct
JWST TRGB--HST SBF overlap and connects to the 63-galaxy HST WFC3/IR SBF sample, giving
$\Hnot=73.8\pm0.7\mathrm{(stat)}\pm2.3\mathrm{(sys)}~\kmsmpc$ \citep{Jensen2025SBF}. Its
dominant systematics are stellar-population color calibration, dust and unresolved sources,
cluster-depth corrections, peculiar velocities, and the zero point of the TRGB or Cepheid distance
scale used to calibrate SBF.

The Tully--Fisher relation connects spiral-galaxy rotation velocity with luminosity, and the
baryonic form replaces luminosity with the sum of stellar and gas mass
\citep{TullyFisher1977,Kourkchi2020,Tully2023}. It provides much larger samples than individual
stellar standard-candle programs: the Cosmicflows-4 Tully--Fisher catalog contains nearly
10,000 spiral galaxies with H\,I line widths and photometry. The statistical power is well matched
for local-flow mapping, bulk-flow studies, and independent $\Hnot$ estimates, but the per-galaxy
scatter is much larger than for SNe Ia or SBF; recent Cosmicflows-4 analyses quote
$\Hnot\simeq75$--76 $\kmsmpc$ with statistical uncertainties of about 2--3 $\kmsmpc$ and
systematic uncertainties of order 1.5--3 $\kmsmpc$ \citep{Kourkchi2020,Tully2023}. The leading
systematics are inclination corrections, internal extinction, linewidth definitions, stellar-mass
calibration in the baryonic relation, Malmquist and selection biases, and the limited number of
level-1 calibrators with Cepheid or TRGB distances.

Type II supernovae replace SNe Ia at the top of the ladder with core-collapse physics. Their
hydrogen-rich spectra give photospheric expansion velocities from broad P-Cygni features; the
standardizable-candle method uses the empirical relation between luminosity and expansion
velocity, with color terms for extinction, while expanding-photosphere and spectral-modeling
methods infer distances more directly from the photospheric radius and flux
\citep{HamuyPinto2002,DeJaeger2022}. Current SNe II samples are smaller and less precise than
SNe Ia samples. The recent local-ladder application used 13 calibrating SNe II with Cepheid or
TRGB host distances and Hubble-flow SNe II to obtain
$\Hnot=75.4^{+3.8}_{-3.7}\mathrm{(stat)}\pm1.5\mathrm{(sys)}~\kmsmpc$, a roughly 5\%
measurement \citep{DeJaeger2022}. Expanding-photosphere measurements avoid an external standard-candle
calibration but depend on dilution factors or radiative-transfer modeling; recent samples of order
10--12 SNe II give uncertainties of several $\kmsmpc$. The dominant uncertainties are velocity
measurements, extinction, explosion diversity, dilution factors or atmosphere models, and the
small number of calibrators observed with sufficient cadence and spectroscopy.

\begin{table}
\begin{center}
\caption{Representative Distance Indicators in the Local Distance Network}
\label{tab:indicators}
\small
\begin{tabular}{L{1.2cm}L{2.0cm}L{2.6cm}L{3.1cm}L{3.1cm}}
\hline\noalign{\smallskip}
Level & Indicator & Population or observable & Calibration & Dominant concerns \\
\hline\noalign{\smallskip}
0 & Gaia parallaxes & Milky Way stars & Direct geometry & Parallax zero point \\
0 & LMC eclipsing binaries & Detached binary stars & External galaxy anchor & Surface-brightness relation \\
0 & NGC 4258 masers & Keplerian maser disk & Direct geometric galaxy distance & Disk modeling \\
1 & Cepheids & Young pulsating stars & Direct SN Ia host calibration & Dust, metallicity, crowding \\
1 & TRGB & Old red giants & $I$-band edge magnitude & Edge detection, AGB contamination \\
1 & JAGB & Carbon-rich AGB stars & NIR luminosity-function mode & Calibration and luminosity-function variations \\
1 & Mira variables & Long-period AGB pulsators & NIR PLR & Circumstellar dust, period cuts \\
1 & RR Lyrae/RRd & Old horizontal-branch pulsators & Optical/NIR PLZ; RRd period ratios & Metallicity, faintness, sample size \\
1 & Contact binaries & W~UMa eclipsing binaries & Period-color-luminosity relation & Binary evolution, third light \\
1 & $\delta$~Scuti/SX Phe & Short-period pulsators & Optical PLR & Mode ID, low luminosity, scatter \\
2 & SNe Ia & Standardized explosions & Width-luminosity-color relation & Color law, host effects, selection \\
2 & SBF & Unresolved stellar fluctuations & Color-dependent fluctuation magnitude & Stellar populations, zero point \\
2 & Tully--Fisher & Rotation-luminosity relation & Linewidth-luminosity relation & Inclination, linewidth, selection \\
2 & SNe II & Expanding-envelope SNe & Velocity-color-light-curve relation & Lower standardization precision \\
\noalign{\smallskip}\hline
\end{tabular}
\end{center}
\end{table}

\begin{figure*}
\centering
\includegraphics[width=0.92\textwidth]{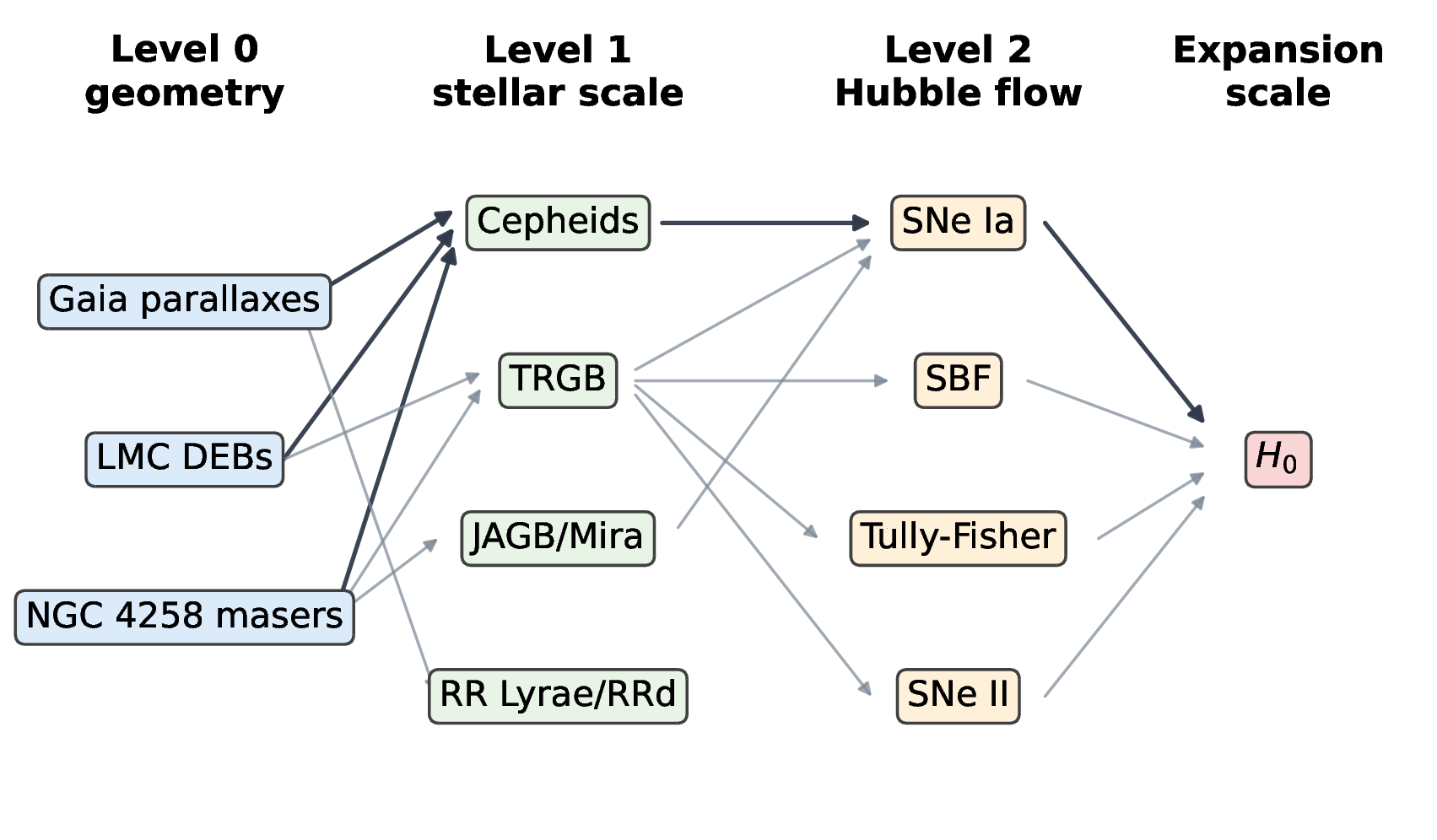}
\caption{A schematic view of the local distance-ladder network used throughout this review.
Level-0 anchors provide geometric distances, level-1 stellar indicators transfer those absolute
scales to nearby galaxies, and level-2 indicators reach the Hubble flow. The network view
emphasizes that level-1 stellar indicators (Cepheids, TRGB, JAGB stars, Miras, and
RR Lyrae/RRd variables) and level-2 observables (SNe Ia, SBF, Tully--Fisher distances, and SNe II) share some calibration
paths but test different astrophysical systematics. Dark arrows mark the canonical
geometric-anchor--Cepheid--SN Ia backbone, while lighter arrows show representative alternative
connections rather than an exhaustive list of all possible calibrations.}
\label{fig:distance_network}
\end{figure*}

Figure~\ref{fig:distance_network} summarizes this bookkeeping as a network of levels and
representative calibration paths.

\section{The Cepheid--SN Ia Distance Ladder}
\label{sect:cepheid}

The Cepheid--SN Ia ladder became the benchmark local route because the HST Key Project compared
Cepheid-calibrated secondary indicators and showed that SNe Ia combine small standardized scatter
with the reach needed to enter the smooth Hubble flow \citep{Freedman2001}. It remains the
highest-precision local route to $\Hnot$. Cepheids work well because their periods and
characteristic light-curve shapes identify them as individual distance indicators, their luminosities
are high enough for observations in nearby SN Ia hosts, and their photometry can be placed on a
common HST system across anchors and calibrator galaxies. The SH0ES strategy deliberately uses
the same instruments and filters across rungs where possible, reducing zero-point discontinuities and
allowing a global fit to Cepheid periods, magnitudes, metallicities, host distances, SN Ia absolute
magnitudes, and the Hubble-flow intercept \citep{Riess2016,Riess2022}.

\subsection{Cepheids as Level-1 Calibrators}
\label{subsect:cepheid_level1}

\begin{figure*}
\centering
\includegraphics[width=0.92\textwidth]{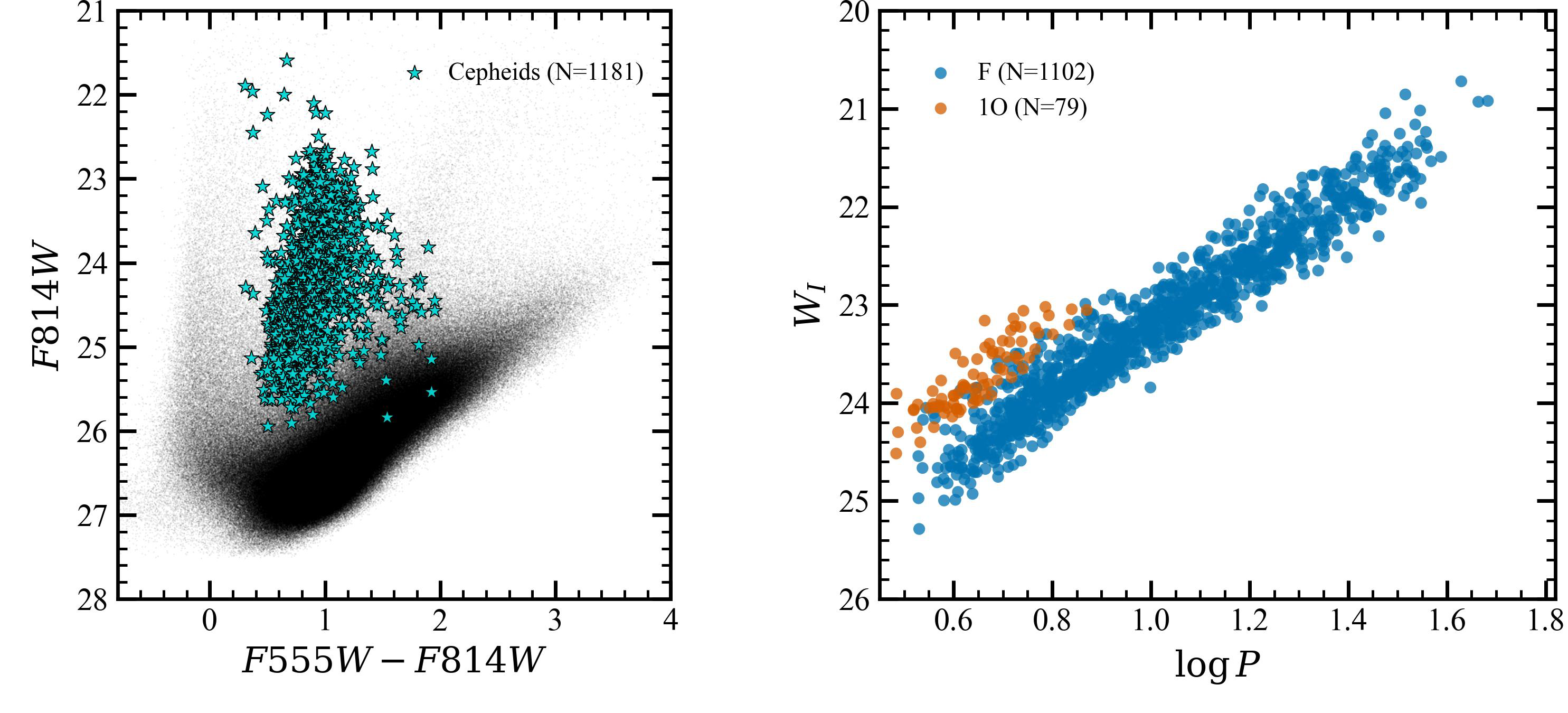}
\caption{Example M101 Cepheid sample in optical color--magnitude and period--Wesenheit space.
Left: the $F555W-F814W$ versus $F814W$ color--magnitude diagram, with matched $F555W$ and
$F814W$ sources shown as rasterized background points and secure Cepheids marked by cyan stars.
Right: the optical Wesenheit period--luminosity relation,
$W_I=F814W-1.3(F555W-F814W)$, for fundamental-mode and first-overtone Cepheids.}
\label{fig:m101_cepheid_cmd_pwr}
\end{figure*}

The Cepheid part of the ladder uses two observational facts at once: the Leavitt law gives the
relative distance scale through periods and magnitudes, and geometric anchors set the absolute
zero point. HST observations of SN Ia hosts generally select long-period Cepheids
($P\gtrsim10$ d) because shorter-period Cepheids fall below the detection limit at
distances of tens of Mpc. The same long-period regime also needs representation in the geometric
anchors, especially in the LMC, NGC 4258, and Milky Way samples, to avoid extrapolating the PLR
from nearby short-period Cepheids to the distant SN Ia host population. The near-infrared Wesenheit magnitude reduces the
effect of dust while preserving sensitivity to the PLR zero point and metallicity term. Operationally,
the Cepheid sample in each host contributes a mean distance modulus, while individual Cepheids
also constrain the common PLR slope, color term, crowding behavior, and metallicity coefficient.
The statistical leverage comes from many stars per galaxy, but the accuracy is controlled
by whether the same PLR describes the Milky Way, LMC, NGC 4258, and SN Ia host samples.
Figure~\ref{fig:m101_cepheid_cmd_pwr} shows a representative M101 example in which Cepheid
selection in the color--magnitude diagram connects directly to the fitted period--Wesenheit
relation.

Before Gaia, HST Fine Guidance Sensor parallaxes provided an early geometric calibration for
10 nearby Galactic Cepheids
\citep{Benedict2007}. HST/WFC3 spatial scanning then pushed accurate astrometry beyond the
traditional sub-kpc regime: the method reaches single-measurement precisions of
$20$--$40~\mu$as and was demonstrated for SY Aur at a distance beyond 2 kpc
\citep{Riess2014SpatialScan}. The first HST spatial-scan Cepheid parallax sample measured seven
long-period ($P>10$ d) Milky Way Cepheids at 1.7--3.6 kpc with a mean precision of
$45~\mu$as and a best precision of $29~\mu$as, increasing the number of long-period Cepheids with
significant direct parallaxes to 10 and giving $\Hnot=73.48\pm1.66~\kmsmpc$ when added to the
2016 ladder \citep{Riess2018Parallax}. A companion HST photometric program put 50 Milky Way
Cepheids onto the same WFC3 photometric system used for extragalactic Cepheids, reducing a
major cross-system uncertainty before Gaia parallaxes were fully folded into the ladder
\citep{Riess2018}.

After the HST Key Project, the precision of the local $\Hnot$ measurement kept improving through
better geometric anchors, infrared Cepheid photometry, and larger samples of Cepheids in SN Ia
host galaxies. The Carnegie Hubble Program used
Spitzer $3.6~\mu$m Cepheid photometry to reduce reddening and metallicity sensitivity: its
calibration used 10 high-metallicity Milky Way Cepheids with trigonometric parallaxes and
80 long-period LMC Cepheids, finding $\Hnot=74.3\pm2.1\mathrm{(sys)}~\kmsmpc$ and a 2.8\%
systematic uncertainty, more than a factor of three smaller than the Key Project systematic error
\citep{Freedman2012CHP}. The 2016 SH0ES analysis then used 19 SN Ia hosts, NGC 4258, the LMC,
and Milky Way Cepheids to obtain a 2.4\% determination,
$\Hnot=73.24\pm1.74~\kmsmpc$ \citep{Riess2016}. The LMC update supplied 70 HST-observed
long-period Cepheids tied to the 1\% detached eclipsing-binary distance; using the LMC alone gave
$\Hnot=74.22\pm1.82~\kmsmpc$, and combining the LMC, NGC 4258, and Milky Way anchors gave
$\Hnot=74.03\pm1.42~\kmsmpc$ \citep{Riess2019}. Gaia EDR3 parallaxes then strengthened the
Milky Way anchor, providing a roughly 1\% Cepheid zero-point calibration while explicitly
marginalizing the bright-Cepheid parallax-offset term \citep{Riess2021}. The 2022 analysis put
these ingredients into a single global solution with 37 Cepheid hosts of 42 SNe Ia and
277 Hubble-flow SNe Ia at $0.0233<z<0.15$ from Pantheon+
\citep{Riess2022,Brout2022}. The same sequence also shows where the statistical bottleneck now sits.
The Hubble-flow rung already contains hundreds of nearby standardized SNe Ia, while the absolute
SN Ia zero point is still tied to a much smaller set of calibrator events and Cepheid host
galaxies. Further statistical improvement now depends mainly on increasing the number of
well-observed nearby SN Ia calibrators and on making their Cepheid distances more homogeneous,
rather than on simply adding more Hubble-flow SNe.

Although this Cepheid ladder reaches sub-percent statistical precision in the global fit,
several systematics can still hide inside the statistical model and require independent validation.
The Cepheid systematic budget begins with the absolute zero point and also includes reddening and
the extinction law, metallicity, photometric cross-calibration, PLR slope and period-distribution
differences, sample completeness, and unresolved background or blending; the last effect is
discussed separately in Section~\ref{subsect:jwst_cepheids}. Completeness remains a difficult
residual systematic because a magnitude-limited Cepheid sample can preferentially lose the
fainter side of the PLR at a given period, shifting the fitted intercept in a way that can correlate
with crowding, background, period, and host distance. This issue is most relevant for the most
distant SN Ia hosts in the current Cepheid sample, including systems with distance moduli near
$\mu\simeq33$ mag ($D\simeq40$ Mpc), where completeness corrections and period cuts are harder
to validate directly.

Gaia EDR3 parallaxes substantially improved the Milky Way Cepheid calibration, but the parallax
zero point remains part of the error model and propagates directly into $\Hnot$ through the
Cepheid luminosity zero point. Using Gaia EDR3 parallaxes together with HST photometry of
75 Milky Way Cepheids, \citet{Riess2021} obtained $\Hnot=73.0\pm1.4~\kmsmpc$ from the
Milky-Way-calibrated Cepheid--SN Ia ladder, and $\Hnot=73.2\pm1.3~\kmsmpc$ when this anchor was
combined with the other geometric calibrators. The Gaia EDR3 bias prescription from the Gaia team
depends on magnitude, color, and sky position \citep{Lindegren2021}; independent checks find
residuals that depend on sample and correction method. \citet{Groenewegen2021} found a mean
quasar-based EDR3 offset of about $-0.021$ mas and an HST-trigonometric-parallax residual of
about $-39~\mu$as before correction, while \citet{Owens2022} concluded that EDR3 bright-Cepheid
parallaxes and Cepheid metallicity together impose a systematic floor of about 3\% for the
Magellanic-Cloud comparison. Recent reanalyses of the Cepheid ladder treat the residual Gaia
EDR3 parallax offset as an explicit nuisance parameter; for example, \citet{Hogas2025} obtain
$zp_{\rm Gaia}=-16\pm6~\mu{\rm as}$ in a full ladder fit and find that the residual-parallax prior
lowers $\Hnot$ by about $0.6~\kmsmpc$ relative to the standard SH0ES treatment, or by about
$1.4~\kmsmpc$ when the Milky Way is used as the sole geometric anchor. \citet{Hogas2026Priors}
adopt a conservative residual-parallax treatment as part of a broader prior reassessment and
obtain a lower combined value, $\Hnot\simeq70.6\pm1.0~\kmsmpc$; this shift is not a
pure Gaia-parallax effect, but it illustrates how parallax zero-point modeling can couple to the
global distance-ladder inference. These studies highlight that the Gaia parallax zero point is now
a model-dependent covariance term; a cleaner external test of this systematic requires the
brighter-star astrometry and revised zero-point calibration expected from Gaia DR4. Metallicity is
similarly unsettled. Multi-galaxy comparisons tied to geometric distances find a moderate negative
term, for example $\gamma_{K_S}=-0.221\pm0.051~{\rm mag~dex^{-1}}$
\citep{Breuval2021Metallicity}, while C-MetaLL analyses based on homogeneous high-resolution
spectroscopic metallicities of Galactic Cepheids find larger coefficients, roughly
$-0.30$ to $-0.55~{\rm mag~dex^{-1}}$ across optical and near-infrared bands
\citep{Bhardwaj2024}. New Gaia releases, JWST photometry, and homogeneous spectroscopic
abundance scales are needed to pin down the metallicity term and to
quantify how different choices of $\gamma$ propagate into the final $\Hnot$ error budget in
global distance-ladder fits.

\subsection{SNe Ia as Level-2 Calibrators}
\label{subsect:shoes_progress}

SNe Ia form the level-2 rung because they connect the nearby galaxy distances measured by
level-1 stellar indicators to the smooth Hubble flow. Their role comes from a physical
regularity across non-identical explosions: thermonuclear disruption of a carbon--oxygen white
dwarf produces a luminous transient, and the peak luminosity can be standardized with
light-curve width, color, and host-galaxy terms
\citep{Phillips1993,Tripp1998,Guy2007,Maoz2014}. In a local distance ladder, the calibration
uses the standardized absolute magnitude $M_B^0$ of the subset that has reliable level-1
distances in the same host galaxies.
The calibrated $M_B^0$ then combines with the Hubble-flow intercept $a_B$ through
Equation~(\ref{eq:h0_formula}).

The past decade mainly increased the calibrator sample and tightened the absolute-magnitude
calibration. The 2016 SH0ES analysis used 19 Cepheid-calibrated SN Ia hosts and obtained a
2.4\% value, $\Hnot=73.24\pm1.74~\kmsmpc$, with a SN Ia absolute magnitude calibrated with the
Spectral Adaptive Lightcurve Template 2 (SALT2) fitter,
near $M_B^0\simeq-19.24$ mag in the primary three-anchor fit \citep{Riess2016}. By 2022 the
Cepheid-calibrated sample had grown to 37 hosts containing 42 SNe Ia, giving
$\Hnot=73.04\pm1.04~\kmsmpc$ and a Pantheon+-standardized absolute magnitude
$M_B^0=-19.253\pm0.027$ mag \citep{Riess2022}. The uncertainty in this absolute magnitude is
about 1.2\% in distance scale, while the calibrator dispersion is 0.130 mag, or about 6\% for a
single standardized SN Ia. The 42 calibrators also provide an internal check on the calibrated
absolute-magnitude distribution: their standardized scatter is nearly the same as the
0.135 mag scatter of the Hubble-flow SN sample, and SH0ES found broad consistency between the
calibrator and Hubble-flow samples in color, light-curve shape, host stellar mass, and
star-formation environment after the baseline cuts and standardization \citep{Riess2022}. The
statistical gain from additional calibrators is real but slow: 42 calibrator SNe remain a
small sample compared with the 1550 SNe Ia in the full Pantheon+ release and the 277 SNe Ia in
the SH0ES baseline Hubble-flow subset, and a few high-leverage objects can matter if they carry
unusual extinction, photometric calibration, host environment, or subtype properties.

The main astrophysical question is whether the calibrator SNe and the Hubble-flow SNe are the
same standardized population after the adopted light-curve, color, and host corrections.
Cosmological analyses exclude strongly peculiar events such as 1991bg-like, Iax, or
super-Chandrasekhar candidates, but the normal SN Ia population still contains spectral and
photometric diversity. 1991T-like events, for example, are luminous slow decliners and can have
different optical and near-infrared absolute-magnitude behavior after standard cuts
\citep{Phillips2022NinetyOneT}. Dust is part of the same matching problem: different extinction
distributions or color--luminosity relations in the calibrator and Hubble-flow hosts can shift the
inferred $M_B^0$ even when each SN sample is internally standardized. The Carnegie Supernova
Project (CSP) analysis uses CSP-I/II optical-to-near-infrared light
curves and combines Cepheid, TRGB, and SBF calibrators in the same CSP Hubble diagram; it
obtains
$\Hnot=71.76\pm0.58\mathrm{(stat)}\pm1.19\mathrm{(sys)}~\kmsmpc$ in the $B$ band and
$\Hnot=73.22\pm0.68\mathrm{(stat)}\pm1.28\mathrm{(sys)}~\kmsmpc$ in the $H$ band, so the
comparison tests passband choice together with the consistency of several level-1 calibrators
\citep{Uddin2024CSP}. The near-infrared SH0ES-style analysis of \citet{Galbany2023NIR} is a
closer test of the Cepheid--SN ladder itself: it uses public $JH$ photometry for up to
19 Cepheid-calibrated SNe Ia from SH0ES hosts and 57 Hubble-flow SNe Ia at $z>0.01$, finding
$\Hnot=72.3\pm1.4\mathrm{(stat)}\pm1.4\mathrm{(sys)}~\kmsmpc$ in $J$ and
$72.3\pm1.3\mathrm{(stat)}\pm1.4\mathrm{(sys)}~\kmsmpc$ in $H$. BayeSN provides a complementary
optical-to-near-infrared hierarchical spectral-energy-distribution (SED) comparison with a fixed
67-SN Hubble-flow sample:
using Cepheid distances to 37 hosts of 41 SNe Ia gives
$\Hnot=74.82\pm0.97\mathrm{(stat)}\pm0.84\mathrm{(sys)}~\kmsmpc$, whereas using TRGB distances
to 15 hosts of 18 SNe Ia gives
$70.92\pm1.14\mathrm{(stat)}\pm1.49\mathrm{(sys)}~\kmsmpc$ \citep{Dhawan2023BayeSN}. The
extinction-model reanalysis of \citet{Wojtak2024Extinction} reweights a SH0ES-like optical
Cepheid-calibrated ladder by applying a consistent host-extinction model to the calibration and
Hubble-flow samples, lowering the result from $73.4\pm1.0$ to $70.5\pm1.0~\kmsmpc$. This
proposal remains debated, but it identifies calibrator--Hubble-flow dust matching as one of the
leading level-2 systematic tests.

The future gain in this rung is set by the supply of well-observed nearby SNe Ia in galaxies where
Cepheids, TRGB, JAGB stars, or Miras can be measured. The realized Cepheid-calibrator sample
grew from 19 SNe Ia in 2016 to 42 in 2022, or roughly four usable calibrators per year over that
interval, but the rate is program-limited because each host needs deep HST or JWST imaging in
addition to high-quality SN photometry and spectroscopy. A simple extrapolation at a comparable
pace would bring the sample to roughly 60--80 calibrator SNe over the next 5--10 yr; coordinated JWST and
HST programs, together with discoveries from the Zwicky Transient Facility (ZTF) and eventually
Rubin/LSST, could push the
calibrator set toward the order of 100 SNe Ia. Purely from the observed
0.130 mag SN calibrator scatter, increasing the calibrator count from 42 to 80--100 would reduce
the calibrator-mean contribution from about 0.9\% to about 0.6--0.7\% in distance. A sub-percent
local $\Hnot$ measurement would require this sample growth to be accompanied by
matching control of dust, subtype selection, photometric cross-calibration, and covariance with
the Hubble-flow sample \citep{Dhawan2022ZTF,Riess2024JWST,Ivezic2019}.

\subsection{Hubble-Flow SNe Ia and the Top-Rung Intercept}
\label{subsect:snia_top}

The Hubble-flow SN Ia sample supplies the final observable needed by the Cepheid--SN Ia
ladder: the intercept $a_B$ of the standardized magnitude--redshift relation. The calibrated
SN Ia absolute magnitude $M_B^0$ fixes the vertical scale of this relation, and
Equation~(\ref{eq:h0_formula}) converts the pair $(M_B^0,a_B)$ into $\Hnot$. At this rung, the
main requirement is a stable zero point for the nearby standardized SN Ia Hubble diagram after
light-curve, host-galaxy, and redshift corrections. In the SH0ES 2022 baseline, this diagram is
fit with 277 Pantheon+ SNe Ia at $0.0233<z<0.15$, restricted to late-type hosts to better match
the Cepheid calibrators \citep{Riess2022,Brout2022}. Its dispersion is 0.135 mag, close to the
0.130 mag dispersion of the calibrator SNe; averaging over 277 objects gives about 0.008 mag in
the mean Hubble-flow level before covariance terms. A coherent 0.01 mag shift in this low-redshift
diagram would move $\Hnot$ by about 0.46\%, or $\simeq0.34~\kmsmpc$ at
$\Hnot\simeq73~\kmsmpc$. The top-rung error budget is controlled by coherent
corrections: survey zero points and filter transformations, SALT2 color--shape and
selection-bias corrections, host-dependent luminosity corrections, and the peculiar-velocity
model.

Pantheon+ currently supplies the baseline low-redshift Hubble diagram used for the
SH0ES/Pantheon+ intercept. The full release contains 1701 light curves of 1550 SNe Ia over
$0.001<z<2.26$ \citep{Scolnic2022}; the local $\Hnot$ measurement is driven by the standardized
nearby Hubble-flow subset and by its covariance with the calibrator hosts. When Pantheon+ is
combined with SH0ES Cepheid host distances, the inferred values are
$H_0=73.4\pm1.1$, $73.5\pm1.1$, and $73.3\pm1.1~\kmsmpc$ for $\Lambda$CDM, flat $w$CDM,
and flat $w_0w_a$CDM, respectively \citep{Brout2022}. Their near equality reflects the
low-redshift nature of the intercept measurement: the cosmological model mainly supplies the
small correction from the linear Hubble law to the luminosity-distance expansion. The
Pantheon+ ingredients most relevant to the top rung are the standardized SN magnitudes,
the survey-by-survey calibration covariance, the bias-correction terms, and the redshift plus
peculiar-velocity covariance used to fit $a_B$ \citep{Brout2022,Peterson2022}.

The first test of $a_B$ is survey calibration. A mixed low-redshift Hubble diagram gains
statistical power, but it also combines different filter systems, photometric zero points, cadence
patterns, and selection functions. The Foundation Supernova Survey reduces this problem by
observing nearby SNe on the Panoramic Survey Telescope and Rapid Response System 1 (Pan-STARRS1)
system; its first release contained 225 SN Ia $griz$
light curves, of which 180 passed the cosmology cuts, and reported an intrinsic scatter of
0.111 mag \citep{Foley2018Foundation,Jones2019Foundation}. The Zwicky Transient Facility
provides an independent homogeneous nearby sample with dense cadence and a well-defined
discovery stream: the first-year release contains 761 spectroscopically classified SNe Ia,
including 305 objects with host-galaxy redshifts suitable for precision cosmology
\citep{Dhawan2022ZTF}. CSP matters here because several TRGB- and CSP-based
$\Hnot$ analyses use CSP photometry for the Hubble-flow rung \citep{Uddin2024CSP}. In the
SH0ES systematic tests, excluding the CSP subset changes the baseline value by only a few tenths
of $\kmsmpc$, while using only CSP for the Hubble-flow sample lowers $\Hnot$ by about
$0.5~\kmsmpc$ \citep{Riess2022}. In the current mixed-survey SH0ES/Pantheon+ construction,
realistic survey calibration errors of 0.025 mag are estimated to contribute only
$\sim0.06~\kmsmpc$ because similar surveys populate the calibrator and Hubble-flow samples; a
single-survey Hubble-flow choice can leave residual survey-mix errors of order $0.8~\kmsmpc$
\citep{Riess2022,Scolnic2022}. The observational goal is a better tied low-redshift sample: duplicate
observations that connect Foundation, ZTF, CSP, and legacy surveys onto the same flux scale,
together with public covariance matrices that show how each survey contributes to $a_B$.

The second test is redshift treatment. The local ladder deliberately uses relatively nearby SNe so
that the inference depends only weakly on the background cosmological model, but the same
choice makes peculiar velocities non-negligible. At the lower edge of the SH0ES Hubble-flow
sample, $z=0.0233$, an uncorrected $250~{\rm km~s^{-1}}$ peculiar velocity corresponds to a
3.6\% velocity perturbation, or about 0.08 mag in distance modulus for a single SN Ia. The
redshift cut, group redshifts, density-field flow corrections, and their covariance are
part of the calibration. In the Pantheon+ redshift analysis, combining group assignments with
coherent-flow corrections lowers the Hubble-residual scatter and raises $\Hnot$ by only
$\sim0.4~\kmsmpc$ relative to using CMB-frame redshifts; the residual method-to-method
uncertainty from redshift corrections is $0.06$--$0.11~\kmsmpc$ in $\Hnot$ \citep{Peterson2022}.
This term is small compared with the Planck--SH0ES difference, but it is already relevant for a
one-percent local measurement and sets the scale required of future flow models.

The third test is population matching between the calibrator SNe and the Hubble-flow SNe after
light-curve, color, dust, and host corrections are applied. In Tripp-style standardization, the
host-galaxy term is commonly parameterized as a step or smooth transition near
$M_\star\simeq10^{10}~M_\odot$, while modern bias simulations include correlations among SN
color, stretch, host mass, dust, and selection \citep{BroutScolnic2021Dust,Brout2022}. SH0ES
mitigates this issue by restricting the baseline Hubble-flow sample to late-type hosts and by
comparing the color, stretch, stellar-mass, and star-formation distributions of the calibrator and
Hubble-flow samples after standardization \citep{Riess2022}. Within this matched
SH0ES/Pantheon+ construction, survey-mixture and flow-model variants move $\Hnot$ at the
few-tenths of $\kmsmpc$ level, far below the $\sim5.7~\kmsmpc$ Planck--SH0ES difference. The
top-rung term that still deserves the strongest stress test is coherent population mismatch,
especially dust, color law, host environment, and selection differences between the calibrator and
Hubble-flow samples; BayeSN and extinction-law reanalyses are well suited to this test even when
their conclusions differ \citep{Dhawan2023BayeSN,Wojtak2024Extinction}. The next benchmark is
an end-to-end analysis in which optical and near-infrared light curves, host-galaxy
properties, group redshifts, peculiar-velocity corrections, and survey calibration terms are varied
together. If such replacements move $\Hnot$ by only $\lesssim0.3$--$0.5~\kmsmpc$, the SN Ia
top rung becomes difficult to regard as the dominant source of the present Hubble tension; if they
move it more, the same tests identify which top-rung term controls the shift.

\subsection{JWST Tests of Cepheid Crowding}
\label{subsect:jwst_cepheids}

The most direct observational concern about the Cepheid route is unresolved near-infrared
crowding in HST images. HST/WFC3/IR has a pixel scale of about $0.13^{\prime\prime}$ pixel$^{-1}$ and a
F160W point-spread-function full width at half maximum (FWHM) of roughly
$0.15^{\prime\prime}$--$0.18^{\prime\prime}$; at $D\simeq40$ Mpc this
corresponds to a physical scale of about 30 pc. Unresolved neighbors and diffuse background
within this scale can make a Cepheid appear brighter and would bias its distance low if left
uncorrected. The SH0ES distance-ladder analyses treat crowding as a measured correction:
artificial stars are injected into the HST images and recovered with the same photometry
pipeline to estimate the crowding correction and its uncertainty in each host environment
\citep{Riess2016,Riess2022}. JWST provides a direct test of this correction because the Near
Infrared Camera (NIRCam) has near-infrared FWHM values of about
$0.05^{\prime\prime}$--$0.07^{\prime\prime}$ in the F150W--F200W range, giving
approximately a factor of three sharper resolution than HST/WFC3/IR at similar wavelengths.

\citet{Riess2024JWST} compared more than 1000 Cepheids in NGC 4258 and five SN Ia hosts
observed with both HST and JWST, finding a mean HST--JWST distance difference of
$-0.01\pm0.03$ mag and rejecting distance-dependent HST crowding as the explanation of the
Hubble tension at 8.2$\sigma$. A later JWST Cycle 2 analysis added a particularly clean test in
NGC 3447A, a low-background SN Ia host environment, and reported no evidence for a Cepheid
photometric bias in the JWST-observed subset of SH0ES hosts \citep{Riess2025PerfectHost}. In that
analysis, 24 SNe Ia in 19 JWST-observed Cepheid hosts gave
$\Hnot=73.49\pm0.93~\kmsmpc$, and the combined Cepheid plus TRGB calibrator set gave
$\Hnot=73.18\pm0.88~\kmsmpc$ \citep{Riess2025PerfectHost}. The remaining tests concern other
systematics; the JWST comparisons provide little support for the specific hypothesis of a large hidden
HST crowding bias.

\section{Independent Distance Ladders and Cross-Checks}
\label{sect:independent}

The Cepheid--SN Ia route currently provides the most precise local value of $\Hnot$, but its
interpretation depends on whether independent distance indicators recover the same galaxy
distances, SN Ia absolute magnitudes, and Hubble-flow intercept. Independent routes are most
informative as diagnostic probes of the distance network. They sample different stellar populations,
wavelength ranges, host environments, and calibration assumptions, and can separate
systematic effects that are difficult to see within a single ladder.

This section follows those routes in the order in which they enter the network. TRGB, JAGB stars,
and Miras are level-1 stellar standard candles that can calibrate nearby SN Ia hosts with stellar
populations different from Cepheids. SBF, Tully--Fisher, and SNe II extend the comparison toward
level-2 or Hubble-flow measurements with different observables and host galaxies. The section then
closes with JWST-era cross-calibration, where the central test is to measure multiple indicators in the
same hosts and to track which zero points, photometric systems, SN samples, and velocity-field
corrections are shared.

\subsection{The TRGB Route}
\label{subsect:trgb}

The TRGB is the most developed independent stellar alternative to Cepheids for calibrating SNe Ia.
Its physical basis is the sharp luminosity cutoff reached by low-mass red giants immediately before
the helium flash. In the optical $I$ band, bolometric and color effects partially compensate, so the
tip magnitude is nearly constant over the old, metal-poor color range normally selected for
precision work \citep{Lee1993,Freedman2020TRGB,Hoyt2023TRGB}. Because these stars can be
measured in galaxy halos, TRGB distances are less exposed to dust, crowding, and young
star-forming structure than Cepheid distances in spiral disks. The method is not assumption-free,
however. Its accuracy depends on how the color-magnitude diagram is selected, how the
luminosity-function edge is detected, how AGB contamination and population mixtures are
controlled, how metallicity or color terms are applied, and how the absolute zero point is tied to
geometric anchors. Empirical multiwavelength calibrations now give TRGB slopes and relative zero
points from $V$ and $I$ through near- and mid-infrared bands, setting the translation from the
classical HST $F814W$ calibration into JWST filters \citep{Madore2023TRGBMW}.

The Carnegie-Chicago Hubble Program (CCHP) used halo TRGB distances for 18 SN Ia
calibrators in 15 host galaxies to calibrate SNe Ia and found
$\Hnot=69.8\pm0.8\mathrm{(stat)}\pm1.7\mathrm{(sys)}~\kmsmpc$ \citep{Freedman2019}. A defining
feature of that result is its zero point: the LMC detached-eclipsing-binary distance gives
$M_I({\rm TRGB})=-4.049\pm0.022\mathrm{(stat)}\pm0.039\mathrm{(sys)}$ mag in the adopted color
range \citep{Freedman2019,Pietrzynski2019}. This value places the corresponding $\Hnot$ between
the Planck \lcdm\ prediction and the Cepheid--SN Ia result, making the Cepheid--TRGB comparison
one of the sharpest internal tests of the local distance scale \citep{Freedman2021}.
Subsequent TRGB studies debate the absolute zero point, the stellar population selected for the
tip, and the mapping from anchor fields to SN Ia hosts. A
SH0ES-side LMC/HST recalibration found a consistent TRGB zero point on the HST photometric
system but a higher SN-calibrated value of $\Hnot$ \citep{Yuan2019TRGB}. The CCHP
multiwavelength recalibration of the LMC, Small Magellanic Cloud (SMC), and IC 1613 obtained an absolute TRGB magnitude,
$M_I({\rm TRGB})=-4.047\pm0.022\mathrm{(stat)}\pm0.039\mathrm{(sys)}$ mag, and then found
$\Hnot=69.6\pm0.8\mathrm{(stat)}\pm1.7\mathrm{(sys)}~\kmsmpc$ \citep{Freedman2020TRGB}. It also
emphasized that the $I$-band TRGB is nearly flat over the old, metal-poor color range used for
precision work. In the Magellanic Clouds, masking young star-forming regions removes
a 0.05--0.10 mag faint bias and gives an old-population zero point
$M_I\simeq-4.050\pm0.030\mathrm{(stat)}\pm0.039\mathrm{(sys)}$ mag
\citep{Hoyt2023TRGB}. The independent NGC 4258 maser anchor gives a closely matching
calibration: \citet{Jang2021TRGB} measure
$F814W_{\rm TRGB}=25.347\pm0.014\mathrm{(stat)}\pm0.042\mathrm{(sys)}$ mag and infer
$M_{\rm F814W}=-4.050\pm0.028\mathrm{(stat)}\pm0.048\mathrm{(sys)}$ mag.

\begin{figure*}
\centering
\includegraphics[width=0.92\textwidth]{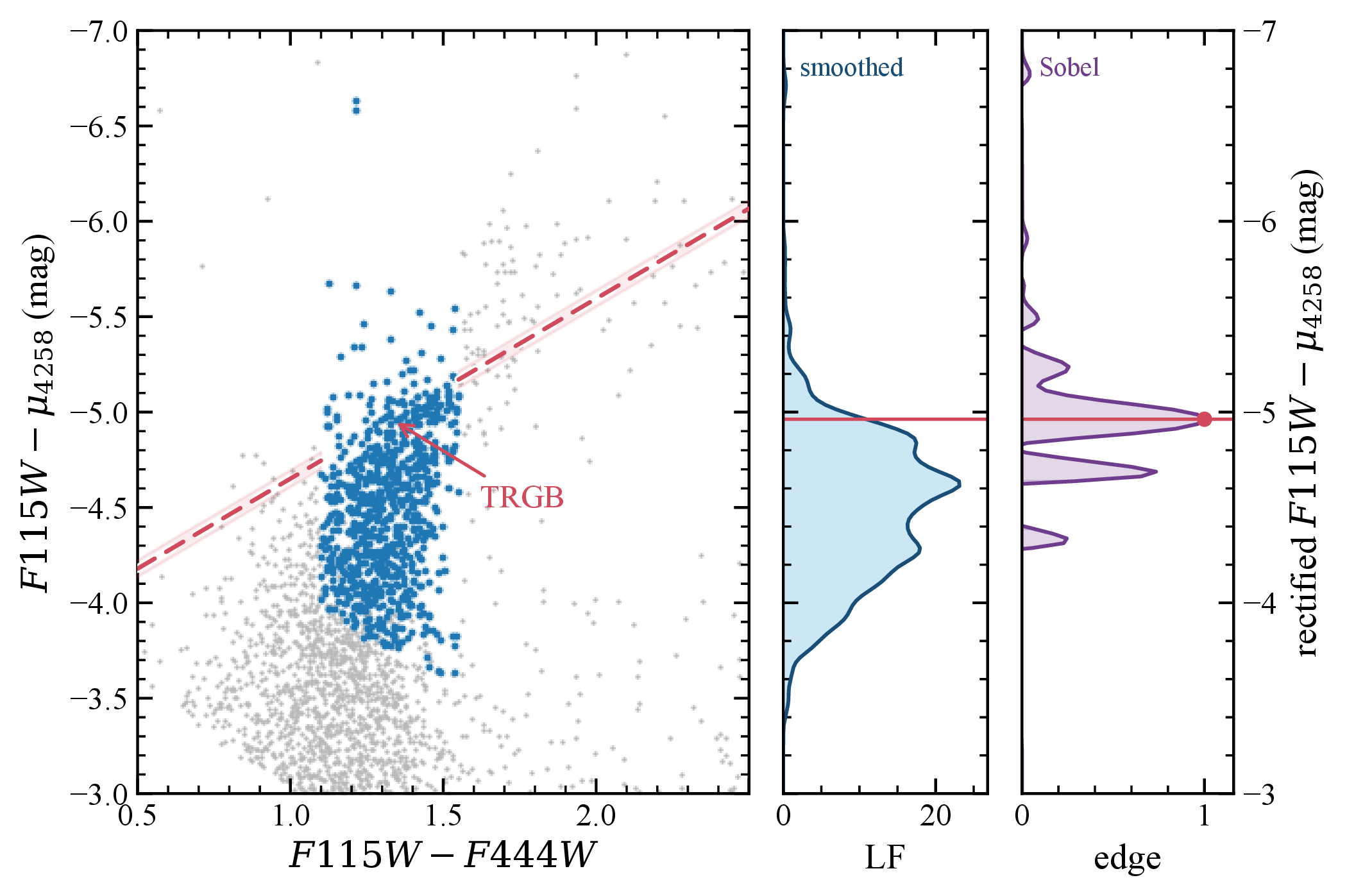}
\caption{Example of anchoring the TRGB luminosity with an edge measurement in an NGC~4258 halo
field observed with JWST.
Left: color--magnitude diagram in $F115W-F444W$ and $F115W-\mu_{4258}$, using
$\mu_{4258}=29.397$ mag. The red line marks the adopted color-corrected TRGB relation used to
rectify the CMD. Middle: smoothed luminosity function in the rectified magnitude. Right:
Sobel edge response, whose peak defines the plotted TRGB location.}
\label{fig:ngc4258_trgb_method}
\end{figure*}

The central measurement problem in TRGB distance work is to make the luminosity-function
edge measured in an anchor field and the edge measured in a SN Ia host field represent the same
standardized old RGB population. A formal edge uncertainty of order 0.01--0.02 mag, as reached in
high-quality anchor-field measurements, is not enough by itself. A single halo field can
show a sharp break while the measured tip still depends on the smoothing scale, color window,
luminosity-function model, and the contrast between red-giant-branch stars and AGB stars above
the tip. These effects are minimized in old, metal-poor halos, but inner fields or mixed stellar
populations can shift the measured edge by several hundredths of a magnitude, for example
0.03--0.05 mag. At the upper end of that range, a 0.05 mag shift corresponds to about a 2.3\%
shift in $\Hnot$. TRGB work has to solve two linked tasks: detecting a statistically sharp
edge and standardizing that edge in the same way in geometric anchors and SN Ia host galaxies.
This motivates artificial-star tests, objective field selection, and direct Cepheid--TRGB
comparisons in the same galaxies.
Figure~\ref{fig:ngc4258_trgb_method} illustrates the same operation in a JWST NGC~4258 halo
field as an example of anchoring the TRGB luminosity to a geometric distance. The
color--magnitude diagram is rectified before constructing the luminosity function and edge
response.

This standardization problem now has a quantitative impact on the $\Hnot$ error budget.
\citet{Anand2022TRGB} compare TRGB distance scales and obtain
$\Hnot\simeq71.5\pm1.8~\kmsmpc$ for one homogeneous calibration route. The CATS standardized-TRGB analysis applies
an automated edge-detection and standardization procedure and reports a baseline value
$\Hnot=73.22\pm2.06~\kmsmpc$; across 108 algorithmic variants, the median is
$72.94\pm1.98~\kmsmpc$ with an additional algorithm uncertainty of $0.83~\kmsmpc$
\citep{Scolnic2023CATS}. A related NGC 4258 field study finds a $\sim0.3$ mag range in
unstandardized $F814W$ tip measurements and introduces a tip-contrast relation with slope
$-0.021\pm0.004$ mag per unit contrast ratio \citep{Li2023TCR}. Together, these papers show that the edge
detector, stellar-population contrast, and choice of anchor-to-host field mapping can move
$\Hnot$ by of order $1$--$2~\kmsmpc$. Simulations of TRGB edge detection likewise show that
photometric errors below about 0.05 mag keep the tip sharp, while errors poorer than about
0.10 mag can introduce false edges and biases \citep{Madore2023TRGBUncert}.

JWST is changing the TRGB discussion by enabling TRGB, Cepheids, and JAGB stars to be observed
in the same SN Ia host galaxies \citep{Hoyt2024,Anand2024JWSTTRGB,Li2024JWSTTRGBII}. JWST
$F090W$ is close to the traditional $I$-band TRGB and retains only a small color term; using
NGC 4258 and two SN Ia hosts, \citet{Anand2024JWSTTRGB} find a mean Cepheid--TRGB distance
difference of $0.01\pm0.06$ mag, too small to explain the Hubble tension. A larger JWST TRGB
sample of eight hosts of 10 SNe Ia similarly finds a weighted-mean TRGB--HST Cepheid difference
of $0.01\pm0.04\mathrm{(stat)}\pm0.04\mathrm{(sys)}$ mag \citep{Li2024JWSTTRGBII}. The CCHP JWST status analysis instead
uses NGC 4258 as the primary geometric anchor, with maser distance
$\mu_{0,4258}=29.397\pm0.024\mathrm{(stat)}\pm0.022\mathrm{(sys)}$ mag, and reports a TRGB-only
best estimate of
$\Hnot=70.39\pm1.22\mathrm{(stat)}\pm1.33\mathrm{(sys)}\pm0.70(\sigma_{\rm SN})~\kmsmpc$
\citep{Reid2019,Freedman2025JWST}. The same analysis gives a JWST $F090W$ TRGB zero point
$M_{\rm F090W}=-4.336\pm0.020\mathrm{(stat)}\pm0.032\mathrm{(sys)}$ mag in the Vega--Sirius
system for NGC 4258 and finds agreement at the level of a few hundredths of a magnitude with
other JWST TRGB calibrations after filter-system offsets are included \citep{Freedman2025JWST}.
In parallel, the CCHP $F115W$ calibration uses a brighter, less-extincted near-infrared tip and
finds repeatability below 0.025 mag among separated NGC 4258 fields, while arguing that some CATS
field choices can be biased at the $>0.4$ mag level \citep{Hoyt2026TRGB}. Because these claims
concern field selection and standardization, they require direct comparison with the CATS
analysis as alternative reductions of overlapping data sets. In its JWST-only subsets, the CCHP
analysis reports
$\Hnot=68.81\pm1.79\mathrm{(stat)}\pm1.32\mathrm{(sys)}~\kmsmpc$ for TRGB and
$\Hnot=67.80\pm2.17\mathrm{(stat)}\pm1.64\mathrm{(sys)}~\kmsmpc$ for JAGB
\citep{Freedman2025JWST}. These values keep TRGB in the center of the debate because the JWST-only
TRGB and JAGB estimates remain about 1.7--1.8$\sigma$ below the Cepheid--SN Ia scale when
statistical and systematic uncertainties are combined. Future improvement requires adding
geometric anchors from the LMC and parallaxes to the present NGC 4258 anchor, and increasing the
number of SN Ia host galaxies with homogeneous TRGB measurements.

\subsection{JAGB Stars and Mira Variables}
\label{subsect:jagb_mira}

JAGB stars are carbon-rich thermally pulsing AGB stars. During third dredge-up, intermediate-age
low- and intermediate-mass stars can reach C/O$>1$ and move into a narrow near-infrared
color--magnitude region; the resulting luminosity function has a well-defined mode that can be
used as a standard candle \citep{MadoreFreedman2020}. The method is usually applied in the
$J$ band or the closely related JWST/NIRCam $F115W$ band, where the stars are bright and the
color term is manageable. Empirically, the JAGB absolute magnitude is near
$M_J\simeq -6.2$ mag, about one magnitude brighter than the near-infrared TRGB. At
$D\simeq40$ Mpc this corresponds to $m_J\simeq26.8$ mag, so JWST can observe JAGB stars in
many present SN Ia hosts with only one or a few near-infrared epochs. Compared with Cepheids,
JAGB stars are older, are selected in lower-surface-brightness outer disks or halos, require no
period recovery, and are less tied to young dusty spiral arms. Their main tradeoff is that the
luminosity-function mode has to be protected against age, metallicity, foreground/background
selection, and field-definition effects.

The current JWST applications quantify both the promise and the remaining calibration problem.
Early observations in NGC 7250, NGC 4536, and NGC 3972 show that JAGB stars are well separated
in color--magnitude space and that outer-disk selection can reduce reddening, blending, and
crowding \citep{Lee2024JAGB}. In their radial-bin procedure, the JAGB luminosity-function mode
stabilizes in the outer regions to within 0.01 mag in NGC 7250, 0.03 mag in NGC 4536, and
0.04 mag in NGC 3972; the observed widths about the individual modes are 0.32, 0.34, and
0.35 mag, respectively, close to the LMC value of about 0.33 mag \citep{Lee2024JAGB}.
\citet{Li2025JAGB} extended the SH0ES-side test to 15 galaxies hosting 18 SNe Ia, calibrated to
NGC 4258, and compared the resulting JAGB distances with HST Cepheid distances. The mean
Cepheid--JAGB offset is
$-0.03\pm0.02\mathrm{(stat)}\pm0.05\mathrm{(sys)}$ mag, so the average scale agrees within the
current errors. The same work also shows why the JAGB zero point is not yet settled: different
ways of measuring the JAGB luminosity function (mode, mean, median, or model) and different
NGC 4258 calibrating fields produce a broad set of possible zero points. Taking the middle value
of this set gives
$\Hnot=73.3\pm1.4\mathrm{(stat)}\pm2.0\mathrm{(sys)}~\kmsmpc$, with the systematic dominated by
the NGC 4258 field dependence; in particular, the two CCHP NGC 4258 fields differ by
$0.11\pm0.022$ mag in the JAGB mode, a shift large enough to move $\Hnot$ by about 5\% if it
entered directly as an absolute-zero-point error. The CCHP JWST JAGB analysis uses its own
blind field-selection algorithm, an NGC 4258 JAGB zero point, and the Carnegie Supernova
Project SN Ia sample; within that framework, seven SN Ia host galaxies give
$\Hnot=67.80\pm2.17\mathrm{(stat)}\pm1.64\mathrm{(sys)}~\kmsmpc$ \citep{Lee2025CCHPJAGB}.
Existing TRGB--JAGB comparisons show inter-method scatter of about 0.07 mag, or roughly 3\%
in distance, indicating that field definition and luminosity-function modeling are now the
limiting pieces of the JAGB route
\citep{Lee2025CCHPJAGB,Li2025JAGB}. JAGB provides an independent check,
with future progress depending on control of field selection, luminosity-function shape, and
anchor-field variance.

Mira variables provide another AGB-based distance measurement, with systematics different from
those of Cepheids, TRGB, and JAGB stars. They are large-amplitude radially pulsating AGB stars,
usually with periods of a few hundred days; the infrared PLR is tight because longer-period Miras
have larger radii and higher luminosities, while observations in $H$, $K$, HST/WFC3 $F160W$, or
mid-infrared bands reduce extinction and temperature sensitivity. Infrared PLRs give scatter of
order 0.12--0.15 mag for suitably selected oxygen-rich Miras and related long-period-variable
(LPV) sequences
\citep{Whitelock2008Mira,Sanders2023MiraGaia,Chen2024LPVMidIR}. At the
$P\simeq240$--$400$ d periods used in current $\Hnot$ work, Miras have typical near-infrared
absolute magnitudes of $M_H$ or $M_K\simeq -7$ to $-8$ mag, about 2--3 mag brighter than a
10 d classical Cepheid in the infrared and comparable to or brighter than the long-period
Cepheids used in SN Ia hosts \citep{Huang2018,Huang2024,Sanders2023MiraGaia}. The
extragalactic Mira ladder was anchored in NGC 4258 by \citet{Huang2018}, who identified
438 Mira candidates and used a 139-object high-quality sample to define an $F160W$ PLR with
about 0.14 mag scatter. The same approach has been applied to SN Ia hosts such as NGC 1559
and M101 \citep{Huang2020,Huang2024}.

Current Mira-based $\Hnot$ measurements are still less precise than the Cepheid--SN Ia or
TRGB--SN Ia routes because only a few SN Ia hosts have sufficiently long time-domain baselines.
The NGC 1559 analysis gave a single-host Mira calibration near
$\Hnot=73.3\pm4.0~\kmsmpc$ \citep{Huang2020}. The M101 analysis used 211 Miras with periods
of 240--400 d and obtained
$\Hnot=72.37\pm2.97~\kmsmpc$, a 4.1\% measurement \citep{Huang2024}. At present the
advantage is primarily diagnostic: Miras probe intermediate-age and old populations, their
variability and period cuts reject contaminants, and their host fields are less restricted to the
young, dusty regions that contain many Cepheids. Their limitations are the
need for multi-year imaging to recover periods, possible population and circumstellar-dust
effects, and smaller current SN Ia host samples.

The observational demands of JAGB stars and Miras are complementary, and both are naturally
suited to JWST and other infrared observations. JAGB distances can be estimated from one or a
few near-infrared epochs once the carbon-star color region is well defined, but the
luminosity-function mode is sensitive to age, metallicity, and field selection. Mira distances
require time-domain imaging over long baselines to recover periods of hundreds of days, but the
PLR can reject contaminants through variability and period cuts. In a multi-indicator JWST
program, the two methods probe intermediate-age and old AGB populations in the same images and
help separate photometric crowding from population-dependent astrophysics.

\subsection{SBF, Tully--Fisher, and Type II Supernovae}
\label{subsect:other}

As alternatives or complements to SNe Ia as level-2 Hubble-flow indicators, several methods can
carry calibrated distances into the Hubble flow without relying on SN Ia luminosity
standardization. The examples most relevant here are SBF, the Tully--Fisher relation, and SNe II.
SBF uses unresolved old stellar populations, Tully--Fisher uses disk-galaxy kinematics over very
large sky areas, and SNe II replace thermonuclear SN Ia standardization with core-collapse physics.
Their value is mainly diagnostic: they test whether the local value of $\Hnot$ depends on young
stellar environments, on the SN Ia top rung, or on local velocity-field modeling.

SBF distances use the finite number of luminous stars in each resolution element of an unresolved
galaxy. The fluctuation magnitude is the ratio of the second and first moments of the
stellar luminosity function, so it is bright in old, metal-rich populations and is calibrated as a
function of integrated color \citep{TonrySchneider1988}. Its natural domain is smooth
early-type galaxies and bulges rather than the star-forming disks used for Cepheid work. This
changes the crowding and dust problem: SBF is less affected by resolving individual stars in
crowded young regions, but its zero point depends on stellar-population modeling, photometric
calibration, and the TRGB or other distances used for the nearby calibrators. The HST WFC3/IR SBF
program measured 63 massive early-type galaxies, including MASSIVE galaxies and SN Ia hosts, out
to about 100 Mpc \citep{Jensen2021SBF,Blakeslee2021}. Its $F110W$ distances have a median
individual uncertainty of about 4\%, an intrinsic calibration scatter near 0.06 mag, and a median
modulus error of 0.083 mag; the resulting direct SBF Hubble-diagram value is
$\Hnot=73.3\pm0.7\mathrm{(stat)}\pm2.4\mathrm{(sys)}~\kmsmpc$ \citep{Blakeslee2021}. A recent
JWST-assisted TRGB--SBF calibration uses eight galaxies with both TRGB and SBF measurements to
recalibrate 61 HST SBF distances, reports a total systematic uncertainty of 0.063 mag, or 2.9\%
in distance, and obtains
$\Hnot=73.8\pm0.7\mathrm{(stat)}\pm2.3\mathrm{(sys)}~\kmsmpc$ \citep{Jensen2025SBF}. In this
implementation SBF avoids Cepheids in the calibration step and does not require SNe Ia as the
top rung; its main shared terms with other low-redshift methods enter through galaxy-group
velocities, local flows, and velocity-field modeling.

The Tully--Fisher relation is based on the empirical connection between the rotation speed of a disk
galaxy and its luminosity or baryonic mass \citep{TullyFisher1977}. Physically, it links the depth
of the dark-matter potential to the stellar and gas content of the disk; observationally, it requires
accurate inclinations, \HI\ linewidths or rotation velocities, photometric corrections, and
selection-function modeling. A single Tully--Fisher distance typically has a modulus uncertainty of
order $0.4$--$0.5$ mag,
corresponding to a distance uncertainty near 20\%, much larger than for a well-observed Cepheid
or TRGB host \citep{Kourkchi2022BTFR,Tully2023}. Its value comes from statistics and sky
coverage. The Cosmicflows-4 compilation contains distances for 55,877 galaxies gathered into
38,065 groups, while its Tully--Fisher catalog provides 9792 spiral-galaxy distances within about
$15,000~\mathrm{km\,s^{-1}}$ \citep{Kourkchi2020TFCatalog,Tully2023}. After redshift and outlier
cuts, recent fits still use thousands of objects, for example 5354 galaxies in a Sloan Digital Sky
Survey (SDSS) $i$-band sample and 3430 galaxies in a Wide-field Infrared Survey Explorer (WISE)
$W1$ sample \citep{Boubel2024TF}. The clearest role of
Tully--Fisher is a large-scale consistency check: it turns the local distance scale into a
quantitative velocity-field test and can reveal whether local flows, bulk motions, or sample
geometry bias the low-redshift Hubble diagram.

As a direct $\Hnot$ route, Tully--Fisher currently gives values on the high side of the local range
with larger systematics than Cepheid--SN Ia. The optical and infrared Cosmicflows-4 calibration
gave a preliminary cluster-based value
$\Hnot=76.0\pm1.1\mathrm{(stat)}\pm2.3\mathrm{(sys)}~\kmsmpc$, and the full Tully--Fisher catalog
gave $\Hnot=75.1\pm0.2\mathrm{(stat)}~\kmsmpc$ with possible systematics up to
$\pm3~\kmsmpc$ \citep{Kourkchi2020}. The baryonic Tully--Fisher relation, calibrated with
64 Cepheid and/or TRGB galaxies and applied to 9984 galaxies extending to roughly $0.05c$,
gave $\Hnot=75.5\pm2.5~\kmsmpc$ \citep{Kourkchi2022BTFR}. More recent Bayesian
forward-modeling of the Tully--Fisher relation and peculiar-velocity field found
$\Hnot=73.3\pm2.1\mathrm{(stat)}\pm3.5\mathrm{(sys)}~\kmsmpc$ in the SDSS $i$ band and
$74.5\pm1.2\mathrm{(stat)}\pm2.6\mathrm{(sys)}~\kmsmpc$ in WISE $W1$; in that framework the
uncertainty from fitting the Tully--Fisher relation itself is only about $0.2~\kmsmpc$, while the
absolute calibrator and zero-point terms dominate \citep{Boubel2024TF}.
\citet{Scolnic2024TF} argued that part of the systematic term in that analysis came from mixing
inconsistent zero-point information, and obtained
$\Hnot=76.3\pm2.1\mathrm{(stat)}\pm1.5\mathrm{(sys)}~\kmsmpc$ using a consistent Cepheid and
TRGB calibration. The method is thus best viewed as an independent late-type-galaxy check on
the local scale and velocity field, with its current precision limited mainly by calibrator zero points,
intrinsic Tully--Fisher scatter, inclination and linewidth systematics, selection effects, and the
peculiar-velocity model.

Type II supernovae can be used as a Hubble-flow distance indicator based on core-collapse
physics. In the expanding-photosphere and standardized-candle approaches, the luminosity is tied
to color, plateau light-curve behavior, and photospheric expansion velocity; brighter events have
larger velocities after correction \citep{HamuyPinto2002}. This physical independence matters because
SNe II progenitors, circumstellar environments, and host-galaxy demographics differ from those of
SNe Ia, while the cost is larger intrinsic diversity and more demanding spectroscopy. Using 13 SNe
II with geometric, Cepheid, or TRGB host distances, \citet{DeJaeger2022} obtained
$\Hnot=75.4^{+3.8}_{-3.7}~\kmsmpc$ with statistical errors only, plus an estimated systematic
uncertainty of $1.5~\kmsmpc$. This is a roughly 5\% measurement rather than a competitor to the
best SN Ia ladders, but it replaces the SN Ia top rung with independent explosion physics and can
become stronger as nearby well-observed SNe II accumulate.

\subsection{Cross-Calibration as the Main JWST-Era Test}
\label{subsect:cross_calibration}

Independent ladders produce separate values of $\Hnot$ and also diagnose hidden
systematics through cross-calibration. The strongest design measures multiple indicators in the
same galaxies, with the same photometric system, the same geometric anchor where possible, and
the same SN Ia calibration when the route uses SNe Ia. Cepheids trace young star-forming regions,
TRGB stars trace old halo populations, and JAGB stars and Miras trace intermediate-age and old AGB
populations. Agreement among these level-1 indicators in the same SN Ia host tests crowding,
reddening, metallicity, field selection, and population-dependent luminosity terms more directly
than comparisons among separate galaxy samples.

JWST turns these pairwise comparisons into a local-distance network. Cepheids, TRGB, JAGB stars,
and Miras can now be measured in overlapping or deliberately matched fields in nearby SN Ia hosts;
SBF, Tully--Fisher, SNe II, masers, and standard sirens add routes with different observables and
host populations
\citep{Riess2024JWST,Anand2024JWSTTRGB,Li2024JWSTTRGBII,Freedman2025JWST,Li2025JAGB,Jensen2025SBF}.
The needed output is a covariance statement as well as a distance comparison. If two routes use
NGC 4258 as the anchor, the same JWST zero point, the same SN Ia light-curve fitter, the same
Pantheon+ Hubble-flow subset, or the same peculiar-velocity model, their agreement mainly tests
the remaining method-specific terms. A difference between two indicators in the same galaxy can
be easier to interpret than a difference between two published values of $\Hnot$, because
the anchor, host, photometric system, and top-rung choices can be held fixed.

Published analyses now put approximate scales on these shared terms. At the SN Ia top rung,
Pantheon+ finds that SN Ia systematics contribute less than one third of the total $\Hnot$
uncertainty \citep{Brout2022}, while redshift and velocity-flow choices within the Pantheon+
Hubble-flow sample leave residual uncertainties of $0.06$--$0.11~\kmsmpc$ in $\Hnot$
\citep{Peterson2022}. At the full-network level, the H0 Distance Network (H0DN) collaboration constructed
a covariance-weighted combination of reviewed local indicators and found a baseline value
$\Hnot=73.50\pm0.81~\kmsmpc$, corresponding to a 1.1\% uncertainty
\citep{H0DN2026}. This baseline is the reference adopted here because it keeps the reviewed
estimators and their covariance assumptions in a controlled network. The combined value is stable
under several leave-one-route tests, and replacing SNe Ia with galaxy-based indicators changes
$\Hnot$ by less than $0.1~\kmsmpc$ while roughly doubling the uncertainty. The methodological
lesson is that agreement among routes is interpreted through their covariance. A one-percent
local distance network requires larger
overlapping samples, improved precision for JAGB stars, Miras, SBF, Tully--Fisher, SNe II, masers,
and standard sirens, and explicit accounting of shared anchors, photometric zero points, SN samples,
and velocity-field corrections.

\section{Systematics, Internal Consistency, and the Hubble Tension}
\label{sect:systematics}

This section recasts the distance-ladder review as an error-budget problem. We first quantify how
small coherent magnitude shifts propagate into $\Hnot$, then separate the level-0 and level-1
terms tied to geometric anchors and stellar indicators, then examine how level-2 indicators and
the Hubble-flow rung reintroduce shared covariance, and finally compare current local values with
the Planck \lcdm\ reference.

\subsection{How Ladder Errors Propagate}
\label{subsect:error_propagation}

The local ladder error budget is easiest to understand by separating three questions: how a
coherent magnitude shift propagates into $\Hnot$, which terms still average down statistically,
and which choices create covariance between otherwise different routes. The propagation can be
seen directly from the distance modulus.
A zero-point shift in a geometric anchor propagates into the level-1 standard-candle calibration,
then into the SN Ia absolute magnitude, and finally into $\Hnot$. A photometric zero point shared
by all Cepheids, a Gaia parallax residual shared by the Galactic anchor, a TRGB edge shift shared
by all halo fields, or a color-law term shared by all SNe Ia behaves very differently from random
scatter that averages down with the number of stars or supernovae. The propagation is logarithmic:
a distance-modulus shift
$\Delta\mu$ corresponds to
\begin{equation}
  \frac{\Delta \Hnot}{\Hnot}\simeq -0.46\,\Delta\mu ,
\end{equation}
with $\Delta\mu$ in magnitudes. Thus 0.01, 0.03, and 0.05 mag zero-point shifts correspond to
approximately 0.5\%, 1.4\%, and 2.3\% shifts in $\Hnot$, respectively. The sign matters:
if the distance-modulus zero point is biased high, the inferred distance scale is too long and
$\Hnot$ is biased low; if it is biased low, $\Hnot$ is biased high. Similarly, a change in the
SN Ia light-curve standardization or Hubble-flow redshift treatment affects all level-1
indicators that use SNe Ia as the top rung.

The statistical and systematic pieces can then be separated. Random scatter in Cepheid PLRs,
SN Ia standardized magnitudes, or Hubble-flow residuals decreases with sample size. Coherent
zero points, population terms, field-selection choices, and shared light-curve or flow corrections
do not. The main remaining statistical bottleneck is the number of SN Ia calibrators in level-1
host galaxies. In the SH0ES 2022 calibration, the standardized calibrator SN dispersion is about
0.130 mag and the sample contains 42 calibrator SNe Ia, so the purely statistical contribution
from the calibrator mean is about 0.020 mag, or about 0.9\% in distance, before covariance terms;
increasing the sample to 80--100 calibrators would reduce this term to roughly 0.013--0.015 mag
\citep{Riess2022}. Most other purely statistical terms are already smaller after averaging many
Cepheids or Hubble-flow SNe. The limiting question for a percent-level $\Hnot$ is
whether the remaining few-hundredths-of-a-magnitude terms are common-mode systematics.

The compact way to keep this bookkeeping honest is a covariance model. The data vector can
include parallaxes, maser or eclipsing-binary distances, Cepheid PLR measurements, TRGB or JAGB
luminosity functions, SN Ia light-curve parameters, redshifts, and external calibration priors.
The parameter vector then contains the anchor distances, level-1 zero points and slopes, SN Ia
absolute magnitude, nuisance parameters, and $\Hnot$. The likelihood depends on a covariance
matrix with both diagonal measurement terms and off-diagonal shared-calibration terms. Adding a
new indicator or a new host improves the final error only to the extent that it breaks one of
these shared terms. Same-host comparisons among Cepheids, TRGB, JAGB stars, and
Miras are more informative than a simple comparison of published $\Hnot$ values: the anchor,
photometric system, host environment, and SN Ia top rung can be held fixed or varied one at a
time.

The scale of the Hubble tension fixes the required size of any hidden offset: moving
$\Hnot=73.04~\kmsmpc$ to the Planck \lcdm\ value $67.36~\kmsmpc$ would require a coherent
distance-modulus change of $5\log_{10}(73.04/67.36)\simeq0.17$ mag, with the sign corresponding
to local distances that would have to become larger or SN Ia absolute magnitudes fainter. This is
much larger than the $-0.01\pm0.03$ mag HST--JWST Cepheid offset measured in the first large
JWST crowding test, but it is comparable to the cumulative effect of several independent
few-hundredths-of-a-magnitude terms if their signs were correlated \citep{Riess2024JWST}. The
central problem is whether small systematics are coherent across anchors, indicators, and the SN Ia
top rung.

A related vulnerability is that source-selection, field-selection, quality-control, and
outlier-rejection choices can enter the covariance model as coherent terms. We therefore treat
AI-assisted, pre-specified selection functions as a future reproducibility requirement rather than
as a separate $\Hnot$ estimator; this workflow is discussed in
Section~\ref{subsect:ai_selection_reproducibility}.

\subsection{Systematics of Level-0 and Level-1 Indicators}
\label{subsect:level1_systematics}

After the propagation scale is fixed, the first systematic for every level-1 indicator is inherited
from the level-0 geometric anchors. The current anchor set is already precise but not exact: the
LMC eclipsing-binary distance is a
1\% distance anchor, the improved NGC 4258 maser distance is a 1.5\% anchor, and the Milky Way
Cepheid parallax scale depends on Gaia parallax zero-point corrections at the few-hundredths of a
magnitude level \citep{Pietrzynski2019,Reid2019,Lindegren2021,Riess2021,Riess2022}. Any
level-1 zero point calibrated on one of these anchors inherits that anchor's coherent error. Adding
more independent anchors reduces the independent part approximately as $1/\sqrt{N_{\rm anchor}}$
and also prevents one geometric system or one calibration field from setting the
entire scale. Common terms, such as Gaia parallax residuals shared by Galactic calibrators or a
systematic difference between the stars used to set an anchor zero point and the stars measured
in SN Ia hosts, are still carried as covariance.

After the inherited anchor term, the remaining systematics are method-specific. Cepheids use
individual young pulsating stars and a PLR or Wesenheit relation; their per-star scatter is only
$0.07$--$0.10$ mag in optimized near-infrared relations, but residual reddening, metallicity,
crowding, incompleteness, and slope choices can produce coherent few-hundredths-of-a-magnitude
shifts \citep{Riess2022,Breuval2022,Riess2024JWST}. TRGB distances use an old-population
luminosity-function edge; formal edge errors can be as small as $0.01$--$0.02$ mag, but color
cuts, smoothing scale, AGB contamination, luminosity-function modeling, and field selection can
move the standardized edge by $0.03$--$0.05$ mag \citep{Madore2023TRGBUncert,Scolnic2023CATS,
Li2024JWSTTRGBII}. JAGB stars use the mode of a carbon-star luminosity function; the raw
luminosity-function width is about $0.32$--$0.35$ mag, and current anchor-field choices can move
the NGC 4258 mode by $0.11\pm0.022$ mag, while TRGB--JAGB comparisons show about
0.07 mag inter-method scatter \citep{Lee2024JAGB,Li2025JAGB,Lee2025CCHPJAGB}. Miras use a
long-period infrared PLR with local scatter of order $0.12$--$0.15$ mag, but their present
extragalactic route is limited by multi-year time baselines, period selection, circumstellar dust,
AGB population differences, and the small number of SN Ia host galaxies
\citep{Huang2018,Huang2024,Sanders2023MiraGaia}.

The way to reduce these terms is not simply to add more stars. Cepheid work benefits most from
near-infrared JWST imaging, artificial-star tests, multi-band reddening constraints, metallicity
leverage, and fixed period and crowding cuts. TRGB work needs objective halo-field selection,
artificial-star recovery, color-window tests, luminosity-function simulations, and direct
Cepheid--TRGB comparisons in the same galaxies. JAGB distances require multiple calibration
fields, multiple anchors, luminosity-function-shape modeling, and blind field-selection rules.
Mira distances require planned time-domain baselines, infrared color and dust diagnostics, and
uniform period-quality cuts. Cross-checks among these level-1 indicators inside the anchor
galaxies are especially diagnostic because the geometric distance is fixed: a disagreement there
isolates the stellar-population or measurement term before the SN Ia top rung is introduced.

New facilities and larger samples are also tests for unknown systematics, not only routes to
smaller formal errors. JWST already checks HST crowding and field selection by changing
wavelength and angular resolution; Roman and Rubin/LSST change the sample selection and
time-domain discovery functions. Future Gaia DR4 and Data Release 5 (DR5) parallaxes, combined
with the LMC
eclipsing-binary distance, can determine whether Galactic and Magellanic Cloud calibrations of
Cepheids, TRGB, JAGB stars, Miras, RR Lyrae stars, and related level-1 indicators are mutually
consistent at the 0.5\% level. That sub-percent cross-anchor test is the scale required for a
one-percent local distance network. Once these level-0 and level-1 terms have been isolated, the
remaining question is how the calibrated distances are carried into the Hubble flow.

\begin{table}
\begin{center}
\caption{Representative Level-0 and Level-1 Systematic Effects}
\label{tab:systematics}
\small
\begin{tabular}{L{2.2cm}L{3.4cm}L{2.8cm}L{3.0cm}}
\hline\noalign{\smallskip}
Term & Examples & Typical scale & Mitigation \\
\hline\noalign{\smallskip}
Geometric anchors & Gaia parallax zero point; LMC eclipsing binaries; NGC 4258 maser model & $\sim1$--1.5\% per major anchor & Multiple independent anchors; shared covariance \\
Cepheids & Reddening, metallicity, crowding, PLR slope, incompleteness & few $0.01$ mag coherent shifts & JWST/NIR imaging; multi-band fits; artificial-star tests \\
TRGB & Edge filter, color window, AGB contamination, halo-field choice & $0.01$--0.02 mag formal edge; $0.03$--0.05 mag field/model shifts & Objective halo selection; simulations; Cepheid--TRGB overlap \\
JAGB & Luminosity-function shape, population mix, anchor-field variance & 0.32--0.35 mag luminosity-function width; up to 0.11 mag field offset & Multiple fields and anchors; blind field rules; luminosity-function modeling \\
Miras & Period recovery, circumstellar dust, AGB population, time baseline & 0.12--0.15 mag local PLR scatter; current $\Hnot$ routes $\sim4$\% & Long-baseline IR monitoring; color cuts; uniform period selection \\
\noalign{\smallskip}\hline
\end{tabular}
\end{center}
\end{table}

\begin{figure}
\centering
\includegraphics[width=0.68\textwidth]{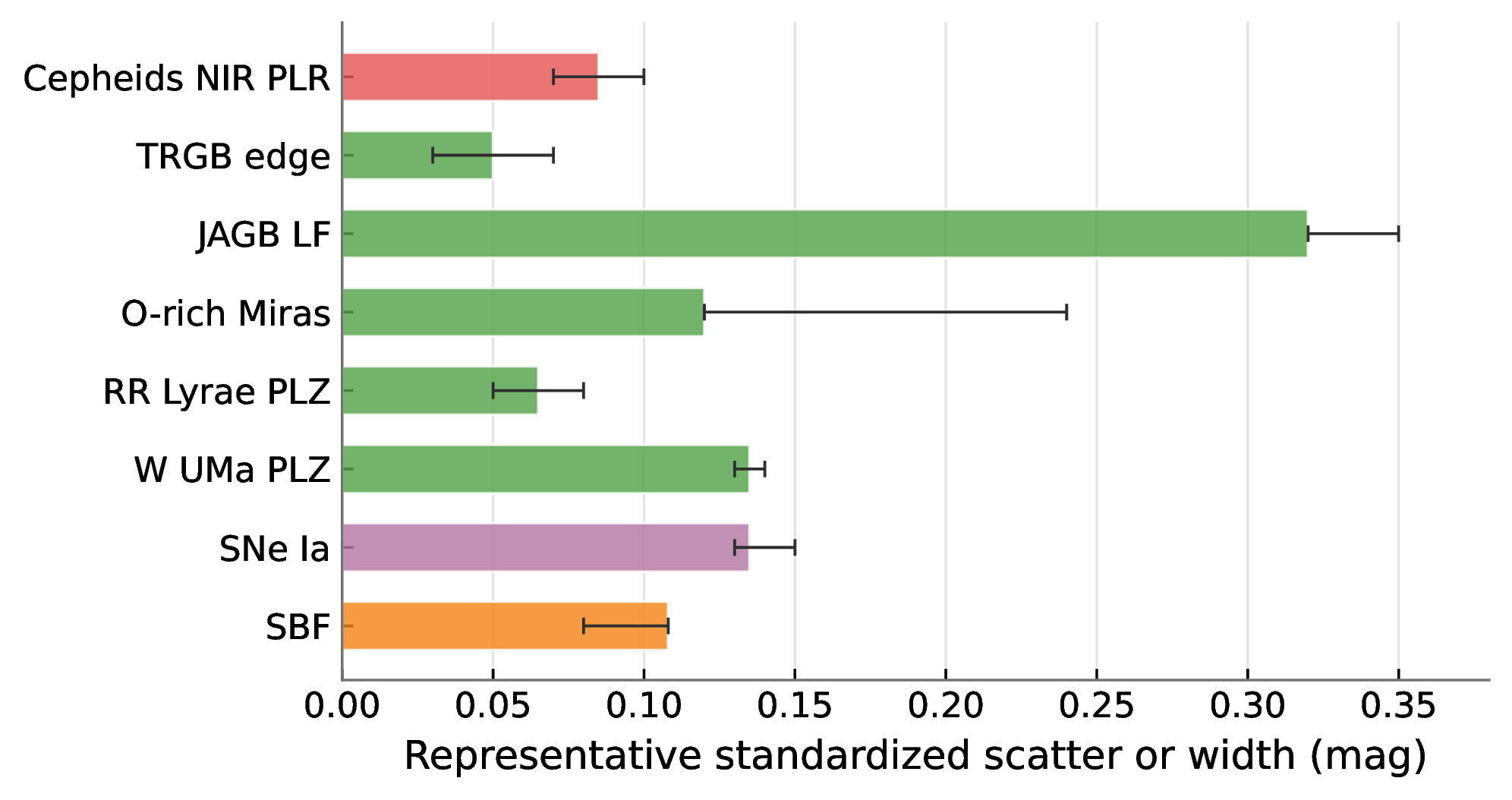}
\caption{Representative scatter or luminosity-function width scales for local distance indicators,
compiled from the recent literature cited in the text. The quantities are not identical observables:
Cepheids, Miras, RR Lyrae, W~UMa systems, and SNe Ia use standardized relations, JAGB uses a
luminosity-function width, TRGB uses an edge measurement, and SBF is shown as an equivalent
magnitude precision for a $\lesssim5\%$ per-galaxy distance. The comparison highlights why large
samples and covariance-aware averaging are needed.}
\label{fig:indicator_scatter}
\end{figure}

Figure~\ref{fig:indicator_scatter} compares representative magnitude scales for several
indicators, emphasizing why per-object scatter and shared systematics must be treated separately.

\subsection{Level-2 Systematics and the Hubble-Flow Rung}
\label{subsect:level2_systematics}

Level-2 indicators are the next error surface in the ladder: they turn calibrated distances in
nearby galaxies into an $\Hnot$ measurement by linking them to objects in the Hubble flow. After
the geometric anchors and level-1 indicators have set the distance scale, three additional
operations matter: the Hubble-flow observable has to be standardized, the calibrator hosts have to
match the Hubble-flow population after that standardization, and the redshifts have to be corrected
onto a common velocity field. For that reason, independent Cepheid, TRGB, JAGB, or Mira calibrations
can become correlated again at the top of the ladder.

For SNe Ia, the statistical Hubble-flow sample is already large. Pantheon+ contains
1701 light curves of 1550 distinct SNe Ia, and the SH0ES baseline intercept uses 277 low-redshift
Hubble-flow SNe Ia selected to reduce peculiar-velocity sensitivity and match the calibrator-host
population \citep{Scolnic2022,Brout2022,Riess2022}. With this sample, redshift and
peculiar-velocity choices in Pantheon+ contribute only about
$0.06$--$0.11~\kmsmpc$ to $\Hnot$ after optimized flow corrections \citep{Peterson2022}. The
larger remaining level-2 tests are tied to SN standardization and calibrator matching. The CSP-I/II
analysis, which fits SNe Ia on its own photometric system and combines Cepheid, TRGB, and SBF
calibrators, quotes systematic terms of $1.19~\kmsmpc$ in $B$ and $1.28~\kmsmpc$ in $H$
\citep{Uddin2024CSP}. BayeSN gives systematic terms of $0.84~\kmsmpc$ for a
Cepheid-calibrated ladder and $1.49~\kmsmpc$ for a TRGB-calibrated ladder
\citep{Dhawan2023BayeSN}. A more extreme extinction-model reanalysis lowers a SH0ES-like
optical result from $73.4\pm1.0$ to $70.5\pm1.0~\kmsmpc$ \citep{Wojtak2024Extinction}. That
proposal remains debated, but it identifies the possible scale of a coherent
calibrator--Hubble-flow dust mismatch.

The matching problem is concrete. Cepheid calibrator hosts are generally star-forming spirals
and preferentially contain nearby, well-observed SNe Ia; TRGB calibrators often use halo fields
in spirals; JAGB and Mira calibrators can select outer-disk or intermediate-age populations. The
Hubble-flow SN Ia sample then mixes host environments, survey selection functions, and
redshift-dependent Malmquist and color-selection effects. The test is whether the
same light-curve or spectral-energy-distribution model gives stable standardized distances across
surveys, passbands, and calibrator populations. Agreement among Pantheon+, CSP, BayeSN, and
near-infrared SN Ia analyses suggests that the high local scale is not driven by one
photometric passband or one light-curve fitter alone, but the residual systematic scale remains of order
$1~\kmsmpc$ for these cross-checks.

Non-SN Ia level-2 routes have different observables but do not yet have comparable precision.
The current JWST TRGB--SBF calibration gives
$\Hnot=73.8\pm0.7\mathrm{(stat)}\pm2.3\mathrm{(sys)}~\kmsmpc$; its limiting terms are the SBF
stellar-population color calibration, group assignment, early-type-galaxy sample selection, and
the inherited TRGB/SBF zero point \citep{Jensen2025SBF}. A consistent Cepheid+TRGB calibration
of the Tully--Fisher relation gives
$76.3\pm2.1\mathrm{(stat)}\pm1.5\mathrm{(sys)}~\kmsmpc$, with the error budget tied to
linewidths, inclination corrections, selection effects, intrinsic relation scatter, and velocity-field
modeling \citep{Scolnic2024TF}. Type II supernovae give
$75.4^{+3.8}_{-3.7}\mathrm{(stat)}\pm1.5\mathrm{(sys)}~\kmsmpc$, where the statistical term is
still dominated by the small calibrated sample and the systematic term includes photospheric
velocity, color/extinction, and plateau-standardization choices \citep{DeJaeger2022}. These
methods are informative because they do not use the same observables as SNe Ia, but their current
uncertainties need to fall to roughly $2~\kmsmpc$ before they can test the Cepheid--SN Ia result at
similar weight.

Thus, the independence of level-2 routes is determined by their covariance, not by their names.
Cepheid--SN Ia, TRGB--SN Ia, JAGB--SN Ia, and Mira--SN Ia routes share the SN Ia Hubble-flow
intercept and light-curve modeling. SBF, Tully--Fisher, and SNe II can share the same
geometric anchors or level-1 calibrator distances, and all low-redshift routes depend on redshift
cuts, group assignments, peculiar-velocity corrections, and flow-model assumptions. Current
comparisons of local $\Hnot$ values have to separate method-specific terms from shared
top-rung terms, which is the bookkeeping used in the method-family comparison and
covariance-weighted summaries below.

\subsection{Current Local Values and the Tension}
\label{subsect:h0_landscape}

\begin{figure*}
\centering
\includegraphics[width=0.88\textwidth]{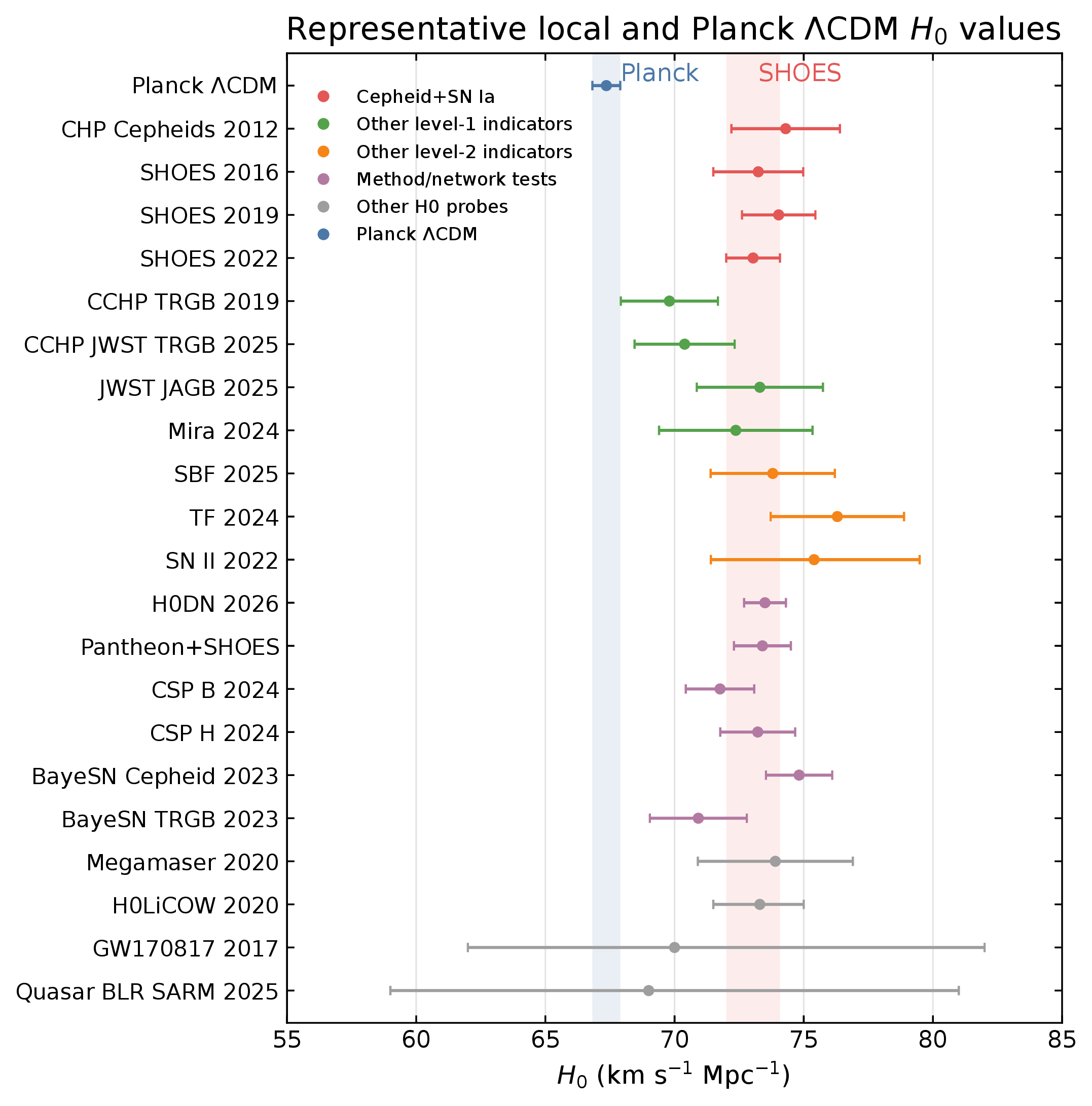}
\caption{Representative values of $\Hnot$ discussed in this review. Statistical and
systematic components are added in quadrature for display where papers quote them separately;
the table in the text preserves the quoted decompositions. The shaded bands show Planck base
\lcdm\ and SH0ES 2022.}
\label{fig:h0_recent}
\end{figure*}

With the level-0, level-1, and level-2 error surfaces separated, the current $\Hnot$ problem
begins with the direct clash between the early-universe and late-universe baselines. Planck 2018
base-\lcdm\ gives
$67.36\pm0.54~\kmsmpc$ (blue point and shaded band in Figure~\ref{fig:h0_recent}), whereas
the Cepheid--SN Ia distance-ladder route gives
$73.04\pm1.04~\kmsmpc$ (red point and shaded band in Figure~\ref{fig:h0_recent})
\citep{Planck2020,Riess2022}. The difference is about
$4.9\sigma$ if the quoted uncertainties are combined in quadrature. This subsection asks
whether distance-ladder variants converge on the Cepheid--SN Ia scale, move toward the Planck
\lcdm\ value, or reveal a specific systematic in one link of the ladder.

Figure~\ref{fig:h0_recent} and Table~\ref{tab:h0_values} summarize representative
determinations. The table is intended as a map of method families and error budgets, since entries
use different anchors, calibrator samples, SN Ia treatments, and covariance assumptions.

\begin{table*}
\begin{center}
\caption{Representative $\Hnot$ Determinations Discussed in This Review}
\label{tab:h0_values}
\scriptsize
\renewcommand{\arraystretch}{0.92}
\setlength{\tabcolsep}{2pt}
\begin{tabular}{L{2.7cm}L{4.2cm}L{3.0cm}L{4.0cm}}
\hline\noalign{\smallskip}
Route & Calibration or data set & $\Hnot$ ($\kmsmpc$) & Main interpretation \\
\hline\noalign{\smallskip}
CMB+\lcdm & Planck 2018 base \lcdm & $67.36\pm0.54$ & Early-universe model-dependent inference \citep{Planck2020} \\
Cepheid--SN Ia & SH0ES 2022, 42 calibrator SNe Ia plus Pantheon+ & $73.04\pm1.04$ & Most precise local ladder baseline \citep{Riess2022} \\
Cepheid--SN Ia & CHP 2012 mid-infrared Cepheids plus SNe Ia & $74.3\pm2.1\mathrm{(sys)}$ & Freedman et al. mid-infrared Cepheid calibration \citep{Freedman2012CHP} \\
SN Ia cosmology & Pantheon+ with SH0ES distances & $73.4\pm1.1$ & SN Hubble-flow covariance and calibration test \citep{Brout2022} \\
TRGB--SN Ia & CCHP HST halo TRGB & $69.8\pm0.8\mathrm{(stat)}\pm1.7\mathrm{(sys)}$ & Lower local route tied to TRGB zero point \citep{Freedman2019} \\
TRGB--SN Ia & CCHP JWST status & $70.39\pm1.22\mathrm{(stat)}\pm1.33\mathrm{(sys)}\pm0.70(\sigma_{\rm SN})$ & NGC 4258 anchored JWST-era comparison \citep{Freedman2025JWST} \\
JAGB--SN Ia & JWST JAGB 2.0 & $73.3\pm1.4\mathrm{(stat)}\pm2.0\mathrm{(sys)}$ & High value with anchor-field systematic \citep{Li2025JAGB} \\
Mira--SN Ia & NGC 4258 Mira PLR plus SN Ia hosts & $72.37\pm2.97$ & Current AGB-pulsator route with two-host-scale precision \citep{Huang2024} \\
SBF & JWST TRGB--SBF calibration & $73.8\pm0.7\mathrm{(stat)}\pm2.3\mathrm{(sys)}$ & Early-type-galaxy route near high scale \citep{Jensen2025SBF} \\
Tully--Fisher & Consistent Cepheid+TRGB calibration of Cosmicflows-4 TF & $76.3\pm2.1\mathrm{(stat)}\pm1.5\mathrm{(sys)}$ & Large-sample late-type-galaxy velocity-field check \citep{Scolnic2024TF} \\
SNe II & Geometric, Cepheid, and TRGB host calibration & $75.4^{+3.8}_{-3.7}\mathrm{(stat)}\pm1.5\mathrm{(sys)}$ & Core-collapse top-rung alternative \citep{DeJaeger2022} \\
SN Ia multi-calibrator & CSP-I/II Cepheid+TRGB+SBF & $71.76\pm0.58\mathrm{(stat)}\pm1.19\mathrm{(sys)}$ in $B$; $73.22\pm0.68\mathrm{(stat)}\pm1.28\mathrm{(sys)}$ in $H$ & Passband and calibrator covariance test \citep{Uddin2024CSP} \\
BayeSN SN Ia & Optical--near-infrared hierarchical SED model & $74.82\pm0.97\mathrm{(stat)}\pm0.84\mathrm{(sys)}$ with Cepheids; $70.92\pm1.14\mathrm{(stat)}\pm1.49\mathrm{(sys)}$ with TRGB & Light-curve model and calibrator dependence \citep{Dhawan2023BayeSN} \\
Local distance network & Covariance-weighted reviewed indicators & $73.50\pm0.81$ & Network consensus with 1.1\% baseline precision \citep{H0DN2026} \\
Seven-route covariance average & Cepheid--SN Ia plus TRGB, JAGB, Mira, SBF, Tully--Fisher, and SNe II & $73.30\pm0.92$ & Fixed-covariance summary of published local routes described in this review \\
\noalign{\smallskip}\hline
\end{tabular}
\end{center}
\end{table*}

Within the distance ladder, the primary empirical line remains the Cepheid--SN Ia route
(red points in Figure~\ref{fig:h0_recent}). It has the largest set of geometric anchors and
SN Ia calibrators, the most mature treatment of Cepheid PLRs, and the
best-tested connection to the Pantheon+ Hubble-flow sample \citep{Riess2022,Brout2022}.
Compared with the already large top-rung Hubble-flow
SN Ia statistics, the next five years can focus on this backbone: refining the
Cepheid PLR in the near-infrared, quantifying selection effects in
extragalactic Cepheid samples, and increasing the number of nearby SN Ia calibrator hosts so
that individual calibrators carry less leverage
\citep{Riess2024JWST,Riess2025PerfectHost}.

The first group of alternatives replaces the Cepheid rung before the SN Ia calibration
(green points in Figure~\ref{fig:h0_recent}). TRGB is the special case because several CCHP
analyses give lower values, from
$69.8\pm0.8\mathrm{(stat)}\pm1.7\mathrm{(sys)}~\kmsmpc$ in the HST halo-field analysis to
$70.39\pm1.22\mathrm{(stat)}\pm1.33\mathrm{(sys)}\pm0.70(\sigma_{\rm SN})~\kmsmpc$ in the
JWST status analysis \citep{Freedman2019,Freedman2025JWST}. This makes TRGB the stellar route
that most clearly needs close study if the local ladder is to move toward the Planck \lcdm\ value.
The tests are geometric rather than cosmetic: applying NGC~4258, LMC, and parallax anchors
in a common framework; comparing anchor fields and SN Ia host fields with the same TRGB
standardization; and completing JWST analyses of larger, homogeneous SN Ia host samples. JAGB
stars and Miras are closer to the high local scale in current SH0ES-side applications, but their
systematics are still dominated by anchor-field choice, luminosity-function definition, period
recovery, dust, and small SN Ia host samples \citep{Li2025JAGB,Huang2024}. They become
competitive only when their total $\Hnot$ uncertainties fall below about
$2~\kmsmpc$.

The second group uses different level-2 or Hubble-flow indicators
(orange points in Figure~\ref{fig:h0_recent}). SBF, Tully--Fisher, and SNe II probe
galaxy populations, velocity fields, and standardization physics that differ from the
Cepheid--SN Ia top rung. Their present central values are generally near the high local scale,
but their total uncertainties remain at roughly $2.4~\kmsmpc$ for SBF,
$2.6~\kmsmpc$ for Tully--Fisher, and about $4~\kmsmpc$ for SNe II
\citep{Jensen2025SBF,Scolnic2024TF,DeJaeger2022}. These routes provide cross-checks, but they
need sharper calibrations, cleaner velocity-field treatment, and larger
homogeneous samples before they can compete directly with the Cepheid--SN Ia precision.

Method-optimization and network analyses play a different role from single-indicator ladders
(purple points in Figure~\ref{fig:h0_recent}). Pantheon+, CSP, BayeSN, and H0DN test how the
answer changes when the SN Ia Hubble-flow sample, passband, light-curve model, calibrator
mixture, and covariance construction are varied
\citep{Brout2022,Uddin2024CSP,Dhawan2023BayeSN,H0DN2026}. They cannot all be counted as
fully independent measurements, because some of them share anchors, SN Ia samples, or
light-curve information. Their results usually remain consistent with the original calibrator or
Hubble-flow samples while providing higher-statistics tests of passband choices, light-curve
models, calibrator mixtures, and covariance assumptions. To make the distance-ladder comparison
easier to interpret, we performed an automated, non-preferential distance-ladder $\Hnot$
calculation with fixed inputs before averaging: the Cepheid--SN Ia route, three
level-1 alternatives that still calibrate SNe Ia (TRGB, JAGB, and Mira), and three level-2
alternatives (SBF, Tully--Fisher, and SNe II).
Table~\ref{tab:seven_route_covariance} lists the dominant systematics in each paper and
separates terms that are not counted as independent. The geometric part is a common scale
component: LMC eclipsing binaries, NGC~4258 masers, and Gaia parallaxes reduce the zero-point
uncertainty when used together, but their residual errors do not become new independent evidence
each time the scale is propagated to another indicator.

\begin{table*}
\begin{center}
\caption{Seven-Route Covariance Model for a Local $\Hnot$ Summary}
\label{tab:seven_route_covariance}
\scriptsize
\renewcommand{\arraystretch}{0.92}
\setlength{\tabcolsep}{2pt}
\begin{tabular}{L{2.6cm}L{2.3cm}L{4.3cm}L{4.8cm}}
\hline\noalign{\smallskip}
Route & Input $\Hnot$ ($\kmsmpc$) & Non-independent terms carried as covariance & Main route-specific systematics \\
\hline\noalign{\smallskip}
Cepheid--SN Ia route & $73.04\pm1.04$ & Shared geometric zero point; SN Ia top-rung intercept & Cepheid crowding, metallicity, color--extinction treatment, photometric zero points, calibrator-host selection \\
TRGB--SN Ia & $70.39\pm1.94$ & NGC~4258/geometric zero point; SN Ia top-rung intercept; overlap with TRGB-calibrated SBF & Edge detection, halo-field selection, RGB/AGB contrast, color window, small SN Ia host sample \\
JAGB--SN Ia & $73.3\pm2.44$ & NGC~4258/geometric zero point; SN Ia top-rung intercept; JWST photometric calibration & JAGB luminosity-function statistic, anchor-field choice, foreground/background selection, carbon-star population dependence \\
Mira--SN Ia & $72.37\pm2.97$ & NGC~4258/geometric zero point; SN Ia top-rung intercept & Period recovery, circumstellar dust, infrared PLR calibration, time-baseline selection, small SN Ia host sample \\
SBF & $73.8\pm2.40$ & TRGB/SBF zero point; local velocity-field terms & Stellar-population color calibration, group assignment, early-type-galaxy sample selection, residual SBF scatter \\
Tully--Fisher & $76.3\pm2.58$ & Cepheid/TRGB calibration inheritance; local velocity-field terms & Linewidths, inclination corrections, Malmquist and selection effects, intrinsic TF scatter, flow modeling \\
SNe II & $75.4\pm4.04$ & Geometric/Cepheid/TRGB host calibration; local velocity-field terms & Photospheric velocity measurements, color and extinction correction, plateau standardization, core-collapse diversity \\
\noalign{\smallskip}\hline
\end{tabular}
\end{center}
\end{table*}

The seven-route combination was computed with a fixed covariance matrix rather than by visually
choosing preferred results. For route values $H_i$ and total quoted uncertainties $\sigma_i$, the
bookkeeping model used
\begin{equation}
C_{ij}=\delta_{ij}\sigma_{i,{\rm ind}}^2+\sigma_{\rm geo}^2
      +I_i^{\rm Ia}I_j^{\rm Ia}\sigma_{\rm Ia}^2
      +I_i^{\rm flow}I_j^{\rm flow}\sigma_{\rm flow}^2 ,
\end{equation}
with $\sigma_{\rm geo}=0.50~\kmsmpc$ common to all seven routes,
$\sigma_{\rm Ia}=0.70~\kmsmpc$ common to the four SN Ia top-rung routes, and
$\sigma_{\rm flow}=0.30~\kmsmpc$ common to SBF, Tully--Fisher, and SNe II. The diagonal
term $\sigma_{i,{\rm ind}}$ is set so that each published total uncertainty is preserved. The
generalized least-squares estimator,
\begin{equation}
\hat H_0=\frac{{\bf 1}^{T}C^{-1}{\bf H}}{{\bf 1}^{T}C^{-1}{\bf 1}},
\qquad
\sigma^2(\hat H_0)=\frac{1}{{\bf 1}^{T}C^{-1}{\bf 1}},
\end{equation}
gives
\begin{equation}
\hat H_{0,{\rm 7r}}=73.30\pm0.92~\kmsmpc ,
\end{equation}
with $\chi^2=4.22$ for 6 degrees of freedom. We performed three internal checks on this result.
A diagonal-only weighted mean gives $73.05\pm0.73~\kmsmpc$, showing that the main effect of
covariance is to widen the uncertainty and move the central value by only $0.25~\kmsmpc$. A
Cholesky Monte Carlo draw from the same covariance matrix reproduces the analytic
$0.92~\kmsmpc$ uncertainty, and leave-one-route-out fits span only
$73.01$--$73.50~\kmsmpc$. The resulting local-ladder value remains $5.6\sigma$ above
Planck base-\lcdm\ when the Planck uncertainty is added in quadrature. It is also very
consistent with the H0DN network summary,
$73.50\pm0.81~\kmsmpc$; the slightly smaller H0DN uncertainty reflects its broader network of
components and covariance constraints.

The resulting local-ladder picture is specific rather than diffuse. Cepheid--SN Ia
remains the primary high-precision route; most non-TRGB alternatives currently agree with the high
local scale within their larger errors; TRGB is the principal lower stellar route and deserves the
most detailed anchor-to-host scrutiny; and network analyses show that covariance-aware
combinations remain near $\Hnot\simeq73~\kmsmpc$. The spread among local routes is small
compared with the historical factor-of-two distance-scale dispute. The caution is that, until other
alternative routes reach comparable precision, possible hidden systematics in the Cepheid--SN Ia
distance ladder still require close scrutiny.

Finally, several late-universe probes provide context outside the classical stellar distance
ladder (gray points in Figure~\ref{fig:h0_recent}). The Megamaser Cosmology Project gives
$\Hnot=73.9\pm3.0~\kmsmpc$ from disk-maser distances \citep{Pesce2020Megamaser}. Strong-lens
time delays gave
$\Hnot=73.3^{+1.7}_{-1.8}~\kmsmpc$ in the H0 Lenses in COSMOGRAIL's Wellspring (H0LiCOW)
six-lens analysis, but the Time-Delay COSMOgraphy (TDCOSMO)
hierarchical treatment showed that galaxy density-profile assumptions can broaden or shift the
lens-inferred value \citep{Wong2020H0LiCOW,Birrer2020TDCOSMO}. This provides a warning:
non-ladder methods can have systematics as subtle as stellar-population or SN Ia calibration
terms. The first binary-neutron-star standard siren, GW170817, yielded
$\Hnot=70.0^{+12.0}_{-8.0}~\kmsmpc$ \citep{Abbott2017StandardSiren}. Quasar
broad-line-region spectroastrometry and reverberation mapping (SARM) provide another geometric route:
the method combines VLTI/GRAVITY angular information with reverberation-mapping linear
sizes, giving $\Hnot=71.5^{+11.9}_{-10.6}~\kmsmpc$ in its first 3C~273 application and
$\Hnot=69^{+12}_{-10}~\kmsmpc$ in a four-quasar analysis \citep{Wang2020BLR,Li2025SARM}. A
separate galaxy-scaling approach uses a proposed universal relation between stellar mass, inferred
from stellar light, and a binding-energy proxy within the effective radius as a type-independent
distance estimator, with a quoted uncertainty of about 0.2 dex in logarithmic distance
\citep{Shi2021BindingEnergy}.
These probes have strong long-term potential, but their current uncertainties or modeling
systematics are larger than those of the best Cepheid--SN Ia and Planck comparisons. Their
near-term role is to serve as external checks on the local-distance network while their samples
and systematic-error models mature; as their samples increase, they are also promising routes to
independent $\Hnot$ determinations.

\section{The JWST Era and Future Prospects}
\label{sect:future}

The next stage of the local-distance program has three concrete tasks. JWST tests whether the
stellar indicators used in the ladder give the same distances when they are measured in the same
galaxies. Gaia, Rubin/LSST, the Chinese Space Station Survey Telescope (CSST), and Roman then
expand the anchors, calibrator hosts, and
time-domain samples needed to reduce statistical bottlenecks. Finally, these data have to enter a
single covariance-aware network so that a smaller error bar also tests for hidden systematics.

\subsection{What JWST Changes}
\label{subsect:jwst_future}

JWST changes the ladder by turning several formerly indirect checks into matched measurements.
Its near-infrared angular resolution lowers crowding and dust sensitivity for Cepheids, and the
same images can also contain RGB and AGB populations for TRGB and JAGB work. The JWST
deliverable is not only a deeper distance to one galaxy; it is a set of Cepheid, TRGB,
JAGB, and eventually Mira measurements in overlapping or deliberately matched fields, tied to the
same geometric anchors and accompanied by the cross-indicator covariance needed for a joint ladder
solution \citep{Riess2024JWST,Riess2024JWSTValidate,Hoyt2024,Freedman2025JWST,
Riess2025PerfectHost}.

For Cepheids, JWST directly tests whether HST crowding has produced a distance-dependent bias.
Current HST--JWST comparisons agree at the few hundredths-of-a-magnitude level and reject hidden
HST crowding as an explanation for the full Hubble tension \citep{Riess2024JWST}. The NGC~3447
``perfect host'' experiment is a cleaner version of the same test: JWST detects about 60
long-period Cepheids in both a normal SN Ia host component and a nearly background-free
star-forming companion at $D\simeq25$ Mpc, finds no component-to-component offset at the
$\lesssim0.03$ mag level, and reduces the Cepheid PLR scatter to about 0.12 mag in the
background-free case \citep{Riess2025PerfectHost}. Across JWST Cycle 1--2 Cepheid observations,
19 SH0ES hosts of 24 SNe Ia have JWST coverage; combining those data with HST gives
$\Hnot=73.49\pm0.93~\kmsmpc$, and adding 35 TRGB-based calibrations gives
$\Hnot=73.18\pm0.88~\kmsmpc$ \citep{Riess2025PerfectHost}. The next Cepheid step is specific:
extend JWST coverage to more SN Ia hosts, improve phase-mean corrections for sparse NIR light
curves, and refit the full HST--JWST calibrator set for an updated Cepheid-based $\Hnot$.

TRGB is the leading JWST alternative because it can be measured in old halo populations of the
same SN Ia host galaxies. The current JWST picture is informative but not yet settled. On the
SH0ES side, NGC~4258-anchored TRGB distances in SN Ia hosts agree with HST Cepheid distances at
the $\sim0.01$ mag level \citep{Anand2024JWSTTRGB,Li2024JWSTTRGBII}. The CCHP JWST program
instead obtains a lower TRGB-based central value,
$\Hnot=70.39\pm1.22\mathrm{(stat)}\pm1.33\mathrm{(sys)}\pm0.70(\sigma_{\rm SN})~\kmsmpc$, and
its JWST-only TRGB subset is lower still, with larger uncertainty \citep{Freedman2025JWST}. A
recent compilation increases the TRGB calibrator sample in normal SN Ia hosts to 35 and shows that,
even on the NGC~4258 anchor, different host subsamples can shift $\Hnot$ because of small-number
statistics and sample selection \citep{Li2026TRGBComplete}. The immediate TRGB tasks are
well defined: more SN Ia hosts, uniform halo-field selection, fixed edge-detection and color-window
rules, JWST--HST filter transformations, and comparison of NGC~4258, LMC, and Gaia-based zero
points \citep{Freedman2020TRGB,Jang2021TRGB,Hoyt2023TRGB,Freedman2025JWST,Li2026TRGBComplete}.

JWST also tests whether newer level-1 indicators introduce their own selection effects. JAGB
measurements probe AGB luminosity functions and field choices, and the NGC~4258 field dependence
shows that a new indicator can reveal new systematics even while providing an independent check
\citep{Hoyt2024,Li2025JAGB}. The near-field program has a complementary role. Nearby dwarf
galaxies and Magellanic-Cloud fields contain old, intermediate-age, and young populations at low
crowding, making them suitable for tying RR Lyrae stars, TRGB, JAGB stars, Miras, and Cepheids
onto one system before exporting those indicators to SN Ia hosts. Gaia-calibrated RR Lyrae work
already places 39 nearby dwarf galaxies on a common distance scale, while Gaia Data Release 3
(DR3) Miras and the
LMC eclipsing-binary distance provide independent anchors for AGB and Cepheid-related ladders
\citep{Pietrzynski2019,Nagarajan2022RRDwarfs,Mullen2023,Sanders2023MiraGaia,
Yan2025JWSTDwarfs,Wang2025JWSTLMCVariables}. The next product is a joint photometric
catalog across Gaia, the LMC, nearby dwarfs, HST, and JWST, using the same artificial-star tests,
crowding diagnostics, metallicity information, and field-selection rules for all level-1
indicators.

\subsection{New Facilities and New Samples}
\label{subsect:future_facilities}

The facilities beyond JWST have a natural division of labor. Gaia improves the geometric
level-0 scale, Rubin/LSST expands the time-domain and calibrator-host samples, CSST supplies a
wide-field HST-independent optical route, and Roman places the enlarged stellar samples onto a
homogeneous near-infrared system.

Gaia DR4 and DR5 primarily target the level-0 scale, because parallax zero-point errors at the
few-microarcsecond level are already relevant for sub-percent distance work. Improved parallaxes for
Galactic Cepheids, RR Lyrae stars, and Mira variables, combined with the LMC
detached-eclipsing-binary distance, can test whether Cepheid, RR Lyrae, Mira,
TRGB, and JAGB zero points are mutually consistent at roughly the 0.5\% level before those level-1
indicators are propagated to SN Ia hosts
\citep{Pietrzynski2019,Lindegren2021,Riess2021,Breuval2022,Mullen2023,Sanders2023MiraGaia}.

Rubin/LSST mainly changes discovery and time-domain sampling. The present SH0ES $\Hnot$ fit uses
37 Cepheid hosts of 42 SNe Ia and about 2150 Cepheids \citep{Riess2022}. Rubin/LSST expands
nearby SNe Ia and SNe II discovery, identifies variable-star populations, and provides of order 100
optical visits in well-sampled fields \citep{Ivezic2019}. This can move the calibrator set from
the current 37 SN Ia host galaxies with Cepheids toward a plausible 60--80 systems and, as a
longer-term target, order 100; it could also increase the usable Cepheid sample toward nearly
$10^4$ stars. The finite-Cepheid statistical
term would then shrink by $\sqrt{2150/10^4}\simeq0.46$, or a factor of about 2.2. At the SN Ia
calibrator rung, increasing the calibrator set from 42 to 80 SNe Ia would reduce the statistical
uncertainty in the SN Ia absolute-magnitude zero point from $0.13/\sqrt{42}=0.020$ mag to
$0.13/\sqrt{80}=0.015$ mag, equivalent to a $0.7\%$ contribution to $\Hnot$ before shared
systematics. These are sample-supply gains; the final one-percent budget still requires the
level-1 zero-point cross-check and covariance accounting discussed below. Rubin/LSST
provides targets and long-baseline classification: it improves variable-star identification,
completeness, and periods, and it tells Roman or JWST where near-infrared follow-up is most
effective.

CSST adds a complementary HST-independent wide-field
route. Its large field of view and planned multi-band imaging and slitless spectroscopic surveys make
it well suited, during its early observing phase, for multi-epoch observations of the Local Group and
nearby SN Ia host galaxies
\citep{Chen2024CSSTVariables,CSSTCollaboration2026}. By observing Cepheids, TRGB, JAGB stars,
Miras, and RR Lyrae stars in the same wide fields across the Local Group and nearby SN Ia host
galaxies, CSST could build a level-1 distance network independent of HST and obtain an independent
$\Hnot$.

Roman's primary role for $\Hnot$ is homogeneous near-infrared resolved-stellar photometry at wide
field. Its imaging provides HST-class angular resolution over larger areas of nearby galaxies and
SN Ia hosts, increasing the number of Cepheids, Miras, and other calibrators in matched host
environments \citep{Spergel2015}. Multi-epoch Roman observations could provide dozens of
near-infrared measurements per star, improving phase-mean magnitude corrections and reducing
sensitivity to crowding and dust, especially for long-period Cepheids in the inner regions of SN Ia
host galaxies. In practice, Roman can take Cepheids and other level-1 indicators identified by
Rubin/LSST and CSST and place them on a common infrared photometric system for an $\Hnot$
measurement.

\subsection{AI-Assisted Selection Functions and Reproducible Distance Ladders}
\label{subsect:ai_selection_reproducibility}

The facility gains above become useful for a one-percent local scale only if they are translated
into homogeneous calibration and reproducible selection. Larger samples reduce Poisson noise and
rare-subclass uncertainty, but they also make cross-survey photometric zero points, field
definitions, light-curve fitting, and velocity-flow modeling more visible sources of covariance.
Future measurement programs therefore need AI-assisted, reproducible workflows that record
pre-specified selection criteria from the start and propagate their effects into the covariance
model, including public artificial-star tests, photometric zero-point chains, light-curve fitter
choices, and redshift-flow models \citep{H0DN2026}.

The need for AI-assisted workflows will become most acute for variable-star distance
indicators, where the path from time-domain photometry to a calibrated distance contains many
coupled selection decisions. For Cepheids, a typical analysis must define source detection and
photometric-quality cuts, variability indices and period-significance thresholds, two-band
light-curve consistency, Fourier-shape constraints, allowed regions in the PL or PW relation,
color and amplitude ranges, crowding and photometric-error limits, final type assignments, and
the rejection of anomalous objects. These thresholds cannot be fixed blindly across all galaxies
because distance, crowding, depth, cadence, stellar background, and star-formation structure vary
from host to host. They must therefore remain physically consistent while also adapting to data
quality and host environment. In the Rubin/LSST, CSST, JWST, and Roman era, the input catalogs can
contain tens of millions to hundreds of millions of sources or detections, so purely manual tuning and
inspection make full reproduction difficult and can reduce the completeness of the recovered
Cepheid, Mira, TRGB, or JAGB samples.

The useful role of AI is procedural and diagnostic. An AI-assisted workflow can collect
the selection logic distributed across source catalogs, light-curve analysis, period searches,
color--magnitude cuts, morphology checks, artificial-star tests, and final distance fits; attach
each threshold to its scientific purpose, pipeline stage, and quantitative diagnostic; and record
how that threshold changes across host galaxies. In this form, AI does not replace the
astrophysical calibration or choose a preferred $\Hnot$. It provides an auditable rule framework
that can standardize terminology, flag inconsistent cuts, identify conflicts between completeness
and reliability, and turn expert judgement into a reusable and reviewable distance-indicator
selection function.

At the distance-ladder level, the same problem becomes statistical rather than merely
procedural. Selection choices determine which stars enter the host distance moduli, which
artificial-star tests define completeness corrections, which outliers are removed, and how
crowding, population differences, photometric zero points, calibrator-host selection, and
SN-calibration terms enter the covariance matrix. A useful AI-assisted analysis should therefore
export survival counts, quality-control diagnostics, completeness and contamination estimates,
decision logs, and covariance terms tied to the frozen selection function. Its purpose is to make
the route from calibrated images and light curves to the final $\Hnot$ likelihood pre-specified,
rerunnable, and externally auditable, with the physical calibration, priors, and final covariance
model kept explicit.

\subsection{Toward a One-Percent Local Scale}
\label{subsect:one_percent}

Recent network analyses show that the field is already close to the one-percent regime. The H0DN
consensus network reaches 1.1\% baseline precision and obtains
$\Hnot=73.50\pm0.81~\kmsmpc$ \citep{H0DN2026}. With larger samples, the main role of such a
network is to check whether different routes continue to converge as the formal errors shrink, or
whether an unrecognized systematic appears in one anchor, one indicator, one SN sample, or one flow
model. Riess et al. argue that Cepheids observed in common with HST and JWST can eventually
calibrate of order 100 SNe Ia and reach a sub-percent local measurement if the photometry and
covariance remain controlled \citep{Riess2024JWST}. The new facilities discussed
above enlarge both the SN calibrator set and the overlapping level-1 indicator
network, allowing convergence tests to be performed as the formal precision improves.

For a credible one-percent local $\Hnot$, these larger samples need to be translated into a
explicit two-rung error budget. The first requirement is a cross-validated level-1 distance-scale
zero point at roughly the 0.5\% level. Gaia DR4/DR5 parallaxes, the LMC
detached-eclipsing-binary distance, NGC 4258, and overlapping JWST/Roman/CSST fields can test
whether Cepheids, TRGB, JAGB stars, Miras, and RR Lyrae stars give mutually consistent distances in
anchors and SN Ia hosts. Agreement at this level would show that the absolute calibration carried
into SN Ia hosts is supported by multiple stellar indicators, multiple anchors, and
pre-specified, reproducible field-selection rules.

The second requirement is the SN Ia absolute-magnitude zero point. With a calibrator scatter of
0.13 mag, increasing the calibrator set from 42 to 80 SNe Ia reduces the statistical uncertainty in
the SN Ia zero point to $0.13/\sqrt{80}=0.015$ mag, or about 0.7\% in $\Hnot$; an order-100
calibrator set would reach about 0.013 mag, or 0.6\%. Combining a 0.5\% level-1 zero point with a
0.7\% SN Ia zero point gives 0.8--0.9\% before peculiar-velocity, photometric, and light-curve
standardization covariance. In practice, reaching below 1\% means realizing these two numbers at
the same time, with the network used to verify that the remaining covariance does not hide a
coherent systematic shift.

\section{Summary}
\label{sect:summary}

The distance ladder has evolved from a single preferred route into a multi-indicator network. The
Cepheid--SN Ia route currently provides the most precise single local ladder and has passed
several major systematic tests, including the specific possibility that HST near-infrared Cepheid
crowding explains the tension. TRGB, JAGB, Mira, SBF, Tully--Fisher, and SNe II routes
are needed because their systematics differ from those of Cepheids and because they can
expose common top-rung effects in SN Ia standardization and Hubble-flow modeling. Recent
covariance-weighted network analyses show that these routes can already be combined at about the
one-percent level. In the fixed seven-route summary used in this review, combining the
Cepheid--SN Ia route with TRGB, JAGB, Mira, SBF, Tully--Fisher, and SNe II routes gives
$\Hnot=73.30\pm0.92~\kmsmpc$, still $5.6\sigma$ above Planck base-\lcdm. With new facilities
producing larger samples of distance indicators, the test is whether Cepheids, TRGB, JAGB
stars, Miras, and other indicators measured in the same galaxies and anchor systems converge to a
common local distance scale under explicit treatment of shared covariance and reproducible,
pre-specified selection criteria. Here the role of AI is to make source
selection, field definition, outlier rejection, quality-control diagnostics, and covariance
construction repeatable before the final $\Hnot$ fit, not to replace the astrophysical calibration
itself. Such convergence is the requirement for a reliable
one-percent local $\Hnot$ measurement. If they do, and the Planck \lcdm\ value remains unchanged,
the case for physics beyond the minimal standard cosmological model becomes stronger. A
disagreement inside the local network at the 0.5\% level would instead identify the
distance-indicator systematic that needs to be solved first.

\begin{acknowledgements}
We thank the anonymous referee for constructive comments that improved the manuscript. We also
thank Dr. Jiyu Wang, Ziming Yan, Pinjian Chen, Prof. Hu Zhan, Prof. Jianming Wang,
Prof. Yong Shi, and all members of the CSST $\Hnot$ project team for helpful discussions and
assistance.
This work was supported by the National Natural Science Foundation of China (NSFC) through grants
12322306, 12373028, and 12173047. This work is supported by the China Manned Space Program with
grant no. CMS-CSST-2025-A01. X. C. and S. W. acknowledge support from the Youth Innovation
Promotion Association of the CAS (grant Nos. 2022055 and 2023065).
\end{acknowledgements}

\bibliographystyle{raa}
\bibliography{references}

\end{document}